\documentclass[12pt]{iopart}
\usepackage{graphicx}
\eqnobysec

\newcommand{\beq}{\begin{equation}}
\newcommand{\beqa}{\begin{eqnarray}}
\newcommand{\eeq}{\end{equation}}
\newcommand{\eeqa}{\end{eqnarray}}
\newcommand{\abs}[1]{\left\vert#1\right\vert}
\newcommand{\bin}[2]{{#1\choose#2}}
\renewcommand{\d}{{\rm d}}
\newcommand{\ds}{\displaystyle}
\renewcommand{\e}{{\rm e}}
\newcommand{\erfc}{\mathop{\rm erfc}\nolimits}
\newcommand{\euler}{{\gamma_{\scriptscriptstyle{\rm E}}}}
\newcommand{\frad}[2]{\ds{\frac{#1}{#2}}}
\newcommand{\fss}{{\rm FSS}}
\renewcommand{\i}{{\rm i}}
\newcommand{\high}{{\rm high}}
\renewcommand{\max}{{\rm max}}
\newcommand{\mean}[1]{\langle#1\rangle}
\renewcommand{\min}{{\rm min}}
\newcommand{\prob}[1]{\mathop{\rm Prob}\nolimits\{#1\}}
\newcommand{\st}{{\rm stat}}
\newcommand{\var}{\mathop{\rm var}\nolimits}
\newcommand{\xic}{\xi_{\rm c}}
\newcommand{\A}{{(\mathrm{A})}}
\newcommand{\B}{{(\mathrm{B})}}
\newcommand{\I}{I_\max}
\newcommand{\K}{k_\max}
\renewcommand{\L}{{\cal L}}
\newcommand{\X}{x_\max}

\begin{document}

\title{Statistics of leaders and lead changes in growing networks}

\author{C Godr\`eche, H Grandclaude and J M Luck}

\address{Institut de Physique Th\'eorique, IPhT, CEA Saclay,
and URA 2306, CNRS, 91191~Gif-sur-Yvette cedex, France}

\begin{abstract}
We investigate various aspects of the statistics of leaders in growing network
models defined by stochastic attachment rules.
The leader is the node with highest degree at a given time
(or the node which reached that degree first if there are co-leaders).
This comprehensive study includes the full distribution
of the degree of the leader, its identity, the number of co-leaders,
as well as several observables characterizing the whole history of lead changes:
number of lead changes, number of distinct leaders,
lead persistence probability.
We successively consider the following network models:
uniform attachment, linear attachment (the Barab\'asi-Albert model),
and generalized preferential attachment with initial attractiveness.
\end{abstract}


\eads{\mailto{claude.godreche@cea.fr},\mailto{helene.grandclaude@cea.fr},\mailto{jean-marc.luck@cea.fr}}

\maketitle

\section{Introduction}
\label{intro}

One of the most salient features of complex networks~\cite{abrmp,doro,blmch,cup}
is their {\it scalefreeness:}
they usually exhibit a broad degree distribution falling off as a power law:
\beq
f_k\sim k^{-\gamma},
\label{f}
\eeq
with $\gamma>2$, so that the mean degree is finite.
Growing networks with preferential attachment,
such as the Barab\'asi-Albert model~\cite{ba},
brought a natural explanation for the ubiquity of the observed scalefreeness.
These model systems also provide a playground
to investigate other, more refined features of networks
and other random growing structures.
In this work we focus our attention onto {\it leaders} and {\it lead changes}.
Luczak and Erd\"os~\cite{el} already describe as {\it a kind of a race}
the growth of connected components
in the Erd\"os-R\'enyi model for random graphs~\cite{er},
and refer to the largest component as the {\it leader}.
This terminology was then introduced in the physics literature
by the pioneering work of Krapivsky and Redner~\cite{lkr},
in concomitance with the growing interest in social networks.
As suggested by these authors, as the degree of a node
may quantify the wealth of a corporation or the popularity of a person,
it is natural to investigate questions such as
{\it How does the identity of the leader change in the course of time?}
{\it What is the probability that a leader
retains the lead as a function of time?}

The first and most natural of all quantities pertaining to this area
is the degree of the leader, i.e., the largest of the node degrees at time $n$.
This quantity is known
to play a central r\^ole in finite-size effects on the degree statistics.
The following picture
has indeed progressively emerged~\cite{dms1,bpv,I}.
For a large but finite scalefree network at time $n$,
the `stationary' power law~(\ref{f}) is cutoff at some time-dependent
scale $k_\star(n)$, which is
of the order of the typical value of the largest degree $\K(n)$.
The latter can in turn be estimated,
at least as a first approximation,
by means of a well-known argument of extreme-value statistics~\cite{evs}:
the stationary probability for the degree $k$
to be larger than $\K(n)$ is of order $1/n$.
In the scalefree case, the largest degree is thus predicted
to grow as a power law:
\beq
\K(n)\sim n^\nu,\quad\nu=\frac{1}{\gamma-1}.
\label{kstar}
\eeq
This growth law is subextensive, as $\gamma>2$ implies $\nu<1$.

The goal of the present paper is to provide a systematic study
of the distribution of the largest degree
and of other quantities related to leaders and lead changes
in growing networks,
thus improving and generalizing most of the results presented in~\cite{lkr,lm}.
We consider network models where a new node enters at each time step,
so that nodes can be identified by an index
equal to the time they enter the network.
We adopt the setup and some of the notations of our earlier work~\cite{I},
devoted to finite-size effects on the degree statistics.
The quantities to be investigated concern either the leader
at a given time or the whole history of lead changes:

\noindent $\bullet$ The {\it largest degree} at time $n$
is by definition the largest of the degrees $k_i(n)$
of all the nodes at time $n$ ($i=1,\dots,n)$:
\beq
\K(n)={\rm max}(k_1(n),\dots,k_n(n)).
\label{kmaxdef}
\eeq
We will investigate both the mean value of $\K(n)$ and its full distribution.

\noindent $\bullet$ A {\it co-leader} at time $n$ is any node
whose degree is equal to the largest degree $\K(n)$.
We denote by $C(n)$ the {\it number of co-leaders} at time $n$.
As the node degrees are integers,
it is not rare that the largest degree (or any other value)
is simultaneously shared by two or more nodes.

\noindent $\bullet$ The {\it leader} at time $n$ is the node
among the co-leaders whose degree reached the value $\K(n)$ {\it first}.
This definition is already used in~\cite{el}.
We denote by~$I(n)$ the {\it index of the leader} at time $n$.
Initial conditions will be such that the first node is the first leader.

\noindent $\bullet$ There is a {\it lead change} at time $n$
if the leader at time $n$ is different from that at time~$n-1$.
We call a {\it lead} any period of time between two lead changes.
We denote by $\L(n)$ the {\it number of leads} up to time $n$.
In other words, the number of lead changes up to time $n$ is $\L(n)-1$.

\noindent $\bullet$ Finally, some lead changes bring to the lead a node
that has already been the leader in the past,
whereas some other changes promote a newcomer.
We denote by $D(n)$ the {\it number of distinct leaders} up to time $n$.
We also introduce the {\it lead persistence probability} $S(n)$
as the probability that there is a single leader up to time $n$.

We will consider network models where
node $n$ attaches to a single earlier node ($i=1,\dots,n-1$),
with a prescribed attachment probability:

\noindent $\bullet$ {\it Uniform attachment} (UA) (Section~2):
the attachment probability is independent of the node,
i.e., uniform over the network.
This model has a geometric degree distribution:
it is therefore not scalefree.
Its analysis however
allows a detailed introduction of key concepts and quantities,
such as the discussion of the `stationary extreme-value statistics'
approach, given in Section 2.1.

\noindent $\bullet$ {\it Barab\'asi-Albert} (BA) {\it model} (Section~3):
the attachment probability is proportional
to the degree of the earlier node.
This well-known model~\cite{ba} is scalefree,
with exponents $\gamma=3$ and $\nu=1/2$.

\noindent $\bullet$ {\it General preferential attachment} (GPA) (Section~4):
the attachment probability is proportional
to the sum $k+c$ of the degree $k$ of the earlier node
and of a constant $c>-1$.
This parameter, representing the initial attractiveness of a node~\cite{dms1},
yields the continuously varying exponents $\gamma=c+3$ and $\nu=1/(c+2)$.

Finally, for each of these models,
it will be interesting to simultaneously analyze data
for two different initial conditions~\cite{I}:

\noindent $\bullet$ {\it Case~A.}
Node 1 appears at time $n=1$ with degree $k_1(1)=0$.
At time $n=2$ node~2 attaches to node~1, so that $k_1(2)=k_2(2)=1$.
At time $n=3$ node~3 attaches a priori either to node~1 or to node~2
with equal probabilities.
We {\it choose} to attach node~3 to node~1,
obtaining thus $k_1(3)=2$, whereas $k_2(3)=k_3(3)=1$,
so that node~1 is the first leader.
This is the initial condition used e.g.~in~\cite{lkr}.

\noindent $\bullet$ {\it Case~B.}
Node 1 appears at time $n=1$ with degree $k_1(1)=1$.
This amounts to saying that the first node is connected to a root,
which does not belong to the network.
It is natural to represent this connection by half a link.
At time $n=2$ node~2 attaches to node~1.
We thus have $k_1(2)=2$ and $k_2(2)=1$,
so that node~1 is again the first leader.

The sum of the degrees $k_i(n)$ of all the nodes at time $n$
equals twice the number of links $L(n)$ in the network~\cite[Eq.~(1.3)]{I}:
\beq
\sum_{i=1}^n k_i(n)=2L(n),
\label{sumr}
\eeq
with $2L^\A(n)=2n-2$ and $2L^\B(n)=2n-1$.
Here and in the following, the super\-scripts~$\A$ and $\B$ denote a result
which holds for a prescribed initial condition (Case~A or Case~B).

\section{The uniform attachment (UA) model}

In the uniform attachment model, each new node $n$
attaches to any earlier node ($i=1,\dots,n-1$)
with uniform probability:
\beq
p_{n,i}=\frac{1}{n-1}.
\eeq

\subsection{Largest degree}

The stationary degree distribution in the UA model has a geometric form:
\beq
f_k=2^{-k}.
\label{fua}
\eeq
It is useful to recall the derivation of this result.
The distribution $f_k(n,i)$ of the degree $k_i(n)$ of node $i$ at time $n$
obeys the recursion~\cite[Eq.~(2.11)]{I}
\beq
nf_k(n+1,i)=f_{k-1}(n,i)+(n-1)f_k(n,i),
\eeq
with $f_k(i,i)=\delta_{k,1}$ for $i\ge2$.
The distribution of the degree of an arbitrary node,
\beq
f_k(n)=\frac{1}{n}\sum_{i=1}^n f_k(n,i),
\label{fkndef}
\eeq
therefore obeys the recursion
\beq
(n+1)f_k(n+1)=f_{k-1}(n)+(n-1)f_k(n)+\delta_{k,1}.
\eeq
Finally, the stationary degree distribution $f_k$
(corresponding to $n\to\infty$ at fixed $k$) obeys the recursion
\beq
2f_k=f_{k-1}+\delta_{k,1},
\eeq
which easily yields~(\ref{fua}).

Applying the extreme-value argument recalled in the Introduction
to the geometric distribution~(\ref{fua}),
we recover the logarithmic growth of the largest degree~\cite{lkr}:
\beq
\K(n)\approx\frac{\ln n}{\ln 2}.
\label{kln}
\eeq
One could be tempted to go further along this line of reasoning
and to assert that the full distribution of the largest degree
$\K(n)$ can be derived, asymptotically for $n$ large,
from the {\it stationary extreme-value statistics} approach.
This consists in modeling $\K(n)$ as the largest of $n$ independent
variables $k_i$ drawn from the stationary distribution $f_k$.
As already suggested in~\cite{lm},
this approach cannot however be exact for the following three reasons,
in order of increasing importance:

\begin{enumerate}

\item[(1)] {\it Sum rule.}
The degrees $k_i(n)$ are not independent random variables,
as they obey the sum rule~(\ref{sumr}).
This condition is however harmless for large $n$,
as it imposes a single constraint on~$n$ variables.

\item[(2)] {\it Finite-size (i.e., finite-time) effects.}
The distribution $f_k(n)$ of the degree of a node at time $n$
does not coincide with its stationary value $f_k$.
It is indeed affected by finite-size effects,
which have been studied in detail in~\cite{I}.
As recalled in the Introduction,
these effects become important for $k\sim k_\star(n)\sim\K(n)$.
Finite-size effects may therefore alter the statistics of the largest degree.

\item[(3)] {\it Correlation between degree and index.}
The degrees $k_i(n)$ of the nodes are not identically distributed.
In particular, older nodes typically have larger degrees.
Their distribution $f_k(n,i)$ may depend rather strongly
on the node index $i$, especially for $i\ll n$.

\end{enumerate}

Points (2) and (3) raise rather delicate issues,
which deserve a specific investigation for each model.
The UA model is a rather fortunate case
where the above points are to a large extent under control~\cite[Sec.~2]{I}.
Concerning point (3),
the dependence of the distribution $f_k(n,i)$ on the index $i$ is rather weak.
We have for instance~\cite[Eq.~(2.4)]{I}
\beq
\mean{k_i(n)}\approx\ln\frac{n}{i}+1
\label{kinmean}
\eeq
for $n$ and $i$ large.
The leading logarithmic growth of this expression
is independent of the node index $i$,
which only enters the `finite part' of the logarithm.
Concerning point~(2), i.e., finite-size effects,
the ratio between the true distribution $f_k(n)$ for $n$ finite
and its stationary counterpart $f_k$ stays approximately equal to unity,
before it drops to zero
in a range of degrees scaling as $(\ln n)^{1/2}$
around $k\approx 2\ln n$.
The prefactor 2 of this estimate is larger than the prefactor
$1/\ln 2\approx1.442695$ of the law~(\ref{kln}),
so that the distribution $f_k(n)$ is expected to be closer and closer
to the stationary one for degrees~$k$ near the mean largest degree.
We are therefore tempted to argue that the
`stationary extreme-value statistics' approach,
or stationary approach for short,
provides a good description of the distribution of the largest degree $\K(n)$
and of related observables,
which may even yield exact asymptotic results,
at least for quantities which are only sensitive
to the leading logarithmic growth of expressions like~(\ref{kinmean}),
and not to their `finite parts'.

The statistics of extremes for i.i.d.~integer variables
is reviewed in Appendix~A, with emphasis on geometric distributions.
The exact distribution of the largest degree is given by~(\ref{phiexact}),
whereas a simpler expression valid for large $n$ is provided by
the `discrete Gumbel law'~(\ref{phiasy}).
The logarithmic growth~(\ref{mvK}) of the mean largest degree
agrees with~(\ref{kln}) for $a=1/2$,
whereas fluctuations are typically of order unity.

We now compare the predictions of the stationary approach
to the outcomes of direct numerical simulations of the model.
Throughout this section on the UA model,
numerical data are obtained by averaging over $10^6$ different histories,
i.e., stochastic realizations of the network,
for both initial conditions (Cases~A and~B),
up to time $n=10^6$, unless otherwise stated.
Statistical errors are smaller (and usually much smaller) than symbols.

\begin{figure}
\begin{center}
\includegraphics[angle=90,width=.45\linewidth]{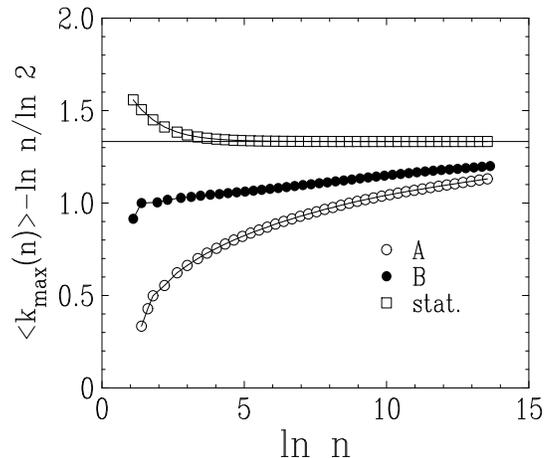}
\caption{\label{figKua}
Plot of the finite part of the mean largest degree,
$\mean{\K(n)}-\ln n/\ln 2$, against $\ln n$.
Circles: numerical data for the UA model with both initial conditions.
Squares and horizontal line: exact stationary values
derived from~(\ref{phiexact}) and their limit (see~(\ref{mvK})).}
\end{center}
\end{figure}

\begin{figure}
\begin{center}
\includegraphics[angle=90,width=.45\linewidth]{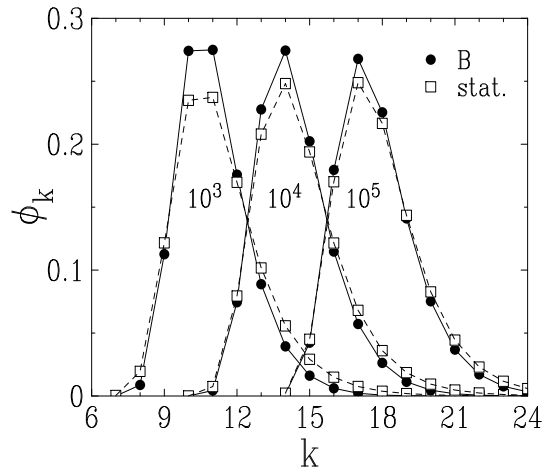}
\caption{\label{figbkua}
Plot of the distribution of the largest degree $\K(n)$
in the UA model for $n=10^3$, $10^4$ and~$10^5$.
Numerical data with initial condition B (circles)
are compared to the exact stationary expression~(\ref{phiexact}) (squares).}
\end{center}
\end{figure}

Figure~\ref{figKua} shows a plot of the finite part
of the mean largest degree, $\mean{\K(n)}-\ln n/\ln 2$,
obtained by subtracting from the data the leading logarithmic law~(\ref{kln}),
against $\ln n$.
The figure also shows the corresponding exact stationary finite part (squares),
derived from~(\ref{phiexact}).
The latter converges very fast to the limit $\euler/\ln 2+1/2\approx1.332746$
(horizontal line),
given by~(\ref{mvK}), where $\euler\approx0.577215$ denotes Euler's constant.
The numerical data (circles) are observed to grow very slowly with $n$.
They however seem to converge to asymptotic values in the same ball park
as the above limit of the stationary finite part.
Figure~\ref{figbkua} shows a plot of the full distribution of $\K(n)$,
for three values of time $n$.
Only data for Case~B are presented, for the sake of clarity.
The exact stationary distributions~(\ref{phiexact}) are plotted
for comparison.
The agreement between numerical data and analytic predictions
of the stationary approach is observed to improve for larger~$n$.
The variability observed in the shape of the distribution
near its top is explained in Appendix~A in terms of periodic oscillations.

\subsection{Number of leads}

The number of leads $\L(n)$ up to time $n$ can be estimated
by elaborating on the observation~\cite{lkr}
that lead changes are driven by the presence of co-leaders.
In the present situation, this line of reasoning yields a quantitative result.
The probability for a lead change to take place at time $n+1$
is indeed proportional to $C(n)-1$,
the number of co-leaders at time $n$ which are not the leader.
Furthermore the $(n+1)$--st node attaches to each of those co-leaders
with probability $1/n$.
We have therefore the exact relation
\beq
\mean{\L(n+1)}-\mean{\L(n)}=\frac{\mean{C(n)}-1}{n}.
\eeq
Within the stationary approach,
the mean number of co-leaders is shown in Appendix~A
to have a finite limit~(\ref{meanC}) for large $n$.
For $a=1/2$ we obtain $\mean{C}=1/\ln 2$.
According to the discussion of Section 2.1,
this leading-order result can be trusted to be asymptotically exact
for large $n$.
Hence
\beq
\mean{\L(n)}\approx A_\L\,\ln n,\quad A_\L=\frac{1}{\ln 2}-1\approx0.442695.
\label{lua}
\eeq

\begin{figure}
\begin{center}
\includegraphics[angle=90,width=.45\linewidth]{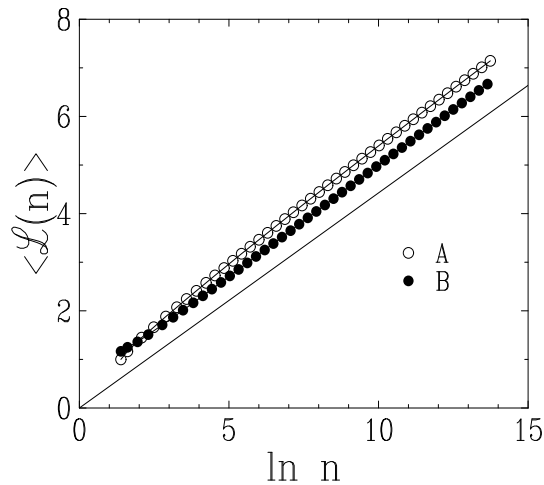}
\caption{\label{figlua}
Plot of the mean number $\mean{\L(n)}$ of leads against $\ln n$,
for the UA model with both initial conditions.
The full line has the theoretical slope $A_\L=1/\ln 2-1\approx0.442695$.}
\end{center}
\end{figure}

Figure~\ref{figlua} shows a plot of $\mean{\L(n)}$ against $\ln n$
for both initial conditions.
The data are observed to follow the logarithmic law~(\ref{lua}),
albeit with appreciable finite-time corrections, especially for Case~A.

\subsection{Index of the leader}

The study of the mean index $\mean{I(n)}$ of the leader at time $n$
also follows the line of reasoning of~\cite{lkr}.
The essential ingredient is the mean index of a node of degree $k$.
This quantity can be derived by the following approach,
which does not require any a priori knowledge of asymptotic degree statistics.
The mean index of a node of degree $k$ at time $n$ reads
\beq
\mean{i_k(n)}=\frac{g_k(n)}{f_k(n)},
\label{meanid}
\eeq
where the denominator $f_k(n)$ has been introduced in~(\ref{fkndef}),
whereas the numerator,
\beq
g_k(n)=\frac{1}{n}\sum_{i=1}^n i\,f_k(n,i),
\label{gkndef}
\eeq
obeys the recursion
\beq
(n+1)g_k(n+1)=g_{k-1}(n)+(n-1)g_k(n)+(n+1)\delta_{k,1}.
\eeq
In the stationary regime ($n\to\infty$ at fixed $k$),
we have $g_k(n)\approx n\gamma_k$, with
\beq
3\gamma_k=\gamma_{k-1}+\delta_{k,1},
\eeq
hence
\beq
\gamma_k=3^{-k}
\eeq
and finally
\beq
\mean{i_k}\approx\left(\frac{2}{3}\right)^kn.
\label{ikua}
\eeq

The mean index of the leader can be estimated by replacing $k$ in~(\ref{ikua})
by the logarithmic law~(\ref{kln}) giving the typical largest degree.
We are thus left with~\cite{lkr}
\beq
\mean{I(n)}\approx A_I\,n^\psi,\quad\psi=2-\frac{\ln 3}{\ln 2}\approx0.415037.
\label{jua}
\eeq
The full-fledged stationary approach allows us to derive a better approximation
of $\mean{I(n)}$, including a numerical estimate for the amplitude.
Within this framework,
averaging the expression~(\ref{ikua}) over the stationary extreme-value
statistics of $\K(n)$ yields $\mean{I(n)}\approx n\,G_n(s)$,
with $s=\ln(2/3)$ and $a=1/2$,
where the generating function $G_n(s)$ is defined in~(\ref{gsum}).
The expressions~(\ref{gres})--(\ref{alphares})
corroborate the power law~(\ref{jua}),
and yield the estimate $A_{I,\st}=\Gamma(\ln3/\ln2-1)/(3\ln2)\approx0.733240$
for the amplitude (with `stat' for `stationary').

\begin{figure}
\begin{center}
\includegraphics[angle=90,width=.45\linewidth]{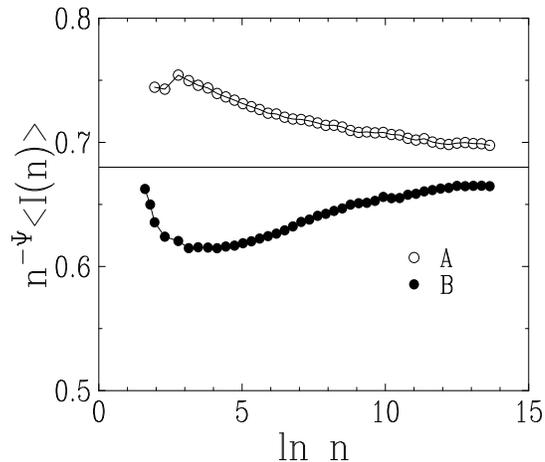}
\caption{\label{figjua}
Plot of the mean index of the leader $\mean{I(n)}$,
for the UA model with both initial conditions,
divided by the theoretical power law $n^\psi$, against $\ln n$.
The horizontal line shows a plausible common limit around $A_I\approx0.68$.}
\end{center}
\end{figure}

Figure~\ref{figjua} shows a plot of the ratio $\mean{I(n)}/n^\psi$
against $\ln n$, for both initial conditions.
Both series of data seem to converge, albeit very slowly, to a common limit
$A_I\approx 0.68$ (full line),
in the same range as the stationary estimate $A_{I,\st}$.

\subsection{Lead persistence probability}

The lead persistence probability $S(n)$ is the probability
that the first node keeps the lead up to time $n$.
It is suggested in~\cite{lkr}
that $S(n)$ can be estimated as the probability
that the degree $k_1(n)$ of the first node
is equal to the mean largest degree, i.e., $\mean{\K(n)}\approx\ln n/\ln 2$
(see~(\ref{kln})), up to negligible finite fluctuation.
Elaborating on this line of thought,
we propose to give a better derivation of $S(n)$
as the probability that the degree $k_1(m)$ of the first node
has been equal to the mean largest degree
for all times $m$ up to the current time $n$.
This improvement takes into account the fact that a persistence phenomenon
is intrinsically non-local in time.

The distribution of the degree $k_1(n)$ of the first node
has been derived in several works.
Consider Case~A for definiteness.
The corresponding generating series reads~\cite{lkr,I}
\beq
F_{n,1}(x)=\sum_{k=1}^\infty f_k(n,1)x^k
=\frac{\Gamma(x+n-1)}{(n-1)!\Gamma(x)},
\eeq
so that we have
\beq
f_k(n,1)=\oint\frac{\d x}{2\pi\i x^{k+1}}\,F_{n,1}(x).
\eeq

In a first step, let us consider the probability for $k_1(n)$
to be very different from its mean value at time~$n$ (see~(\ref{kinmean})),
namely $k_1(n)\approx b\ln n$ with $b\ne1$.
This probability can be estimated by applying the saddle-point approximation
to the above contour integral.
We thus obtain a logarithmic large-deviation estimate of the form
\beq
f_{b\ln n}(n,1)\sim(\ln n)^{-1/2}\,n^{-\Phi(b)},\quad\Phi(b)=1-b+b\ln b,
\eeq
with $\Phi(b)\approx(b-1)^2/2$ as $b\to1$.
The above estimate holds irrespective of the initial condition.
In the case of interest, i.e., $b=1/\ln 2$,
the exponent reads $\phi=\Phi(1/\ln 2)\approx0.086071$~\cite{lkr}.
In a second step, we estimate the lead persistence probability
as the probability that $k_1(m)$ has been of order $b\ln m$
for all times $m$ up to time $n$, and not only at the current time $n$.
The effect of such a condition has been investigated in detail
in the case of the one-dimensional lattice random walk~\cite{drunk}.
As a general rule, the mere effect
is to change the exponent of the `prefactor' from $1/2$ to $3/2$.
We are therefore left with the estimate
\beq
S(n)\sim(\ln n)^{-3/2}\,n^{-\phi}.
\label{suares}
\eeq

\begin{figure}
\begin{center}
\includegraphics[angle=90,width=.45\linewidth]{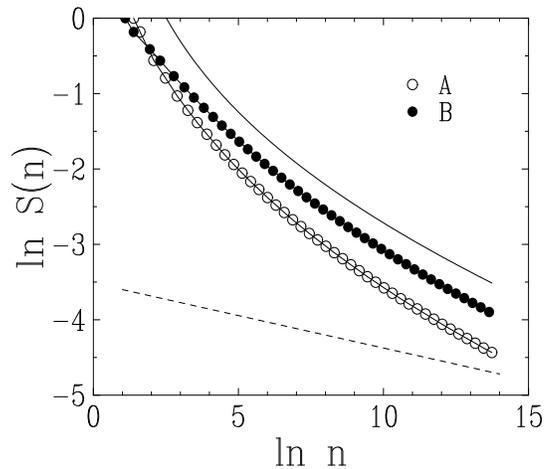}
\caption{\label{figsua}
Log-log plot of the lead persistence probability $S(n)$ against time $n$,
Symbols: Data for the UA model with both initial conditions.
Dashed and full line:
asymptotic power law $n^{-\phi}$ and full estimate~(\ref{suares}),
with arbitrary numerical prefactors chosen for readability.}
\end{center}
\end{figure}

Figure~\ref{figsua} shows a log-log plot of the lead
persistence probability
against time $n$, for both initial conditions and $n$ up to $10^6$.
The data points are far from being aligned,
indicating thus a strong deviation from a pure power law~\cite{lkr}.
Moreover, the effective slope over the plotted data range is around 0.3,
i.e., four times larger than the theoretical asymptotic exponent $\phi$
(dashed line).
The full estimate~(\ref{suares}) (full line),
including the prefactor $(\ln n)^{-3/2}$,
gives a much better representation of the data.

\subsection{Number of distinct leaders}

The mean number of distinct leaders up to time $n$
is expected to grow logarithmically:
\beq
\mean{D(n)}\approx A_D\,\ln n.
\label{meandua}
\eeq
Two heuristic reasons can be given in favor of this law.
First, the number of lead changes up to time $n$, $\L(n)\sim\ln n$,
grows much less rapidly than the number of candidates to the lead,
which can be estimated as $I(n)\sim n^\psi$.
It is therefore likely that any lead change brings a newcomer to the lead,
with some non-zero probability $\Pi$.
Second, the growth of the largest degree $\K(n)\sim\ln n$ is modest,
so that newcomers have `a short way to go' before they can reach the lead.

\begin{figure}
\begin{center}
\includegraphics[angle=90,width=.45\linewidth]{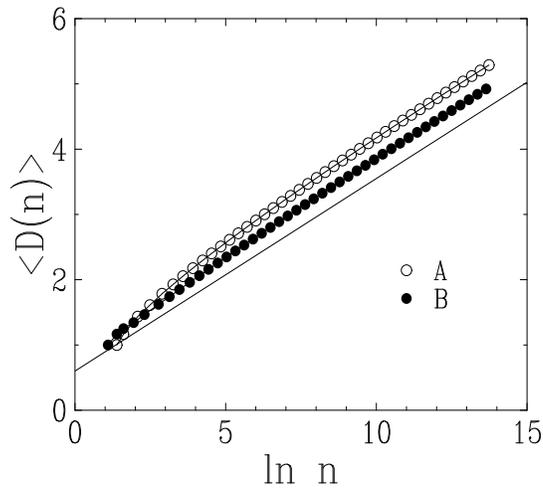}
\caption{\label{figdua}
Plot of the mean number $\mean{D(n)}$ of distinct leaders up to time $n$,
against time $n$, for the UA model with both initial conditions.
The full line has slope $A_D\approx0.29$.}
\end{center}
\end{figure}

Figure~\ref{figdua} shows a plot of the mean number $\mean{D(n)}$
of distinct leaders, against time~$n$, for both initial conditions.
The picture is somewhat similar to that of Figure~\ref{figlua},
with appreciable finite-time corrections, especially for Case~A.
The data confirm our expectation,
namely an asymptotic logarithmic law of the form~(\ref{meandua}),
irrespective of the initial condition.
The slope $A_D\approx0.29$ of the full line is the outcome of a fit including
a correction in $1/n$.
The ratio of $A_D$ to the amplitude $A_\L$ of the growth law~(\ref{lua})
of the mean number of~leads
yields the probability for any lead change to bring a newcomer to the lead:
\beq
\Pi=\frac{A_D}{A_\L}\approx0.65.
\eeq

\section{Linear preferential attachment: the Barab\'asi-Albert (BA) model}

In the Barab\'asi-Albert (BA) model,
node $n$ attaches to any earlier node
with a probability proportional to the degree of that node.
The attachment probability of node $n$ to node $i$ thus reads
\beq
p_{n,i}=\frac{k_i(n-1)}{Z(n-1)},
\label{pba}
\eeq
where the partition function in the denominator ensures the normalization
of the probabilities.
It reads $Z(n)=2L(n)$ (see~(\ref{sumr})), i.e., $Z^\A(n)=2n-2$, $Z^\B(n)=2n-1$.

\subsection{Largest degree}

The stationary degree distribution of the BA model
can be derived along the lines of Section~2.1.
The distribution $f_k(n,i)$ of the degree of node $i$ at time $n$
obeys the recursion
\beq
Z(n)f_k(n+1,i)=(k-1)f_{k-1}(n,i)+(Z(n)-k)f_k(n,i),
\eeq
with $f_k(i,i)=\delta_{k,1}$ for $i\ge2$.
The distribution of the degree of an arbitrary node
therefore obeys the recursion
\beq
(n+1)Z(n)f_k(n+1)=n(k-1)f_{k-1}(n)+n(Z(n)-k)f_k(n)+\delta_{k,1}.
\eeq
Finally, the stationary degree distribution $f_k$
on an infinite network obeys the recursion
\beq
(k+2)f_k=(k-1)f_{k-1}+2\delta_{k,1},
\eeq
hence
\beq
f_k=\frac{4}{k(k+1)(k+2)}.
\label{fba}
\eeq

For a large degree $k$,
the stationary distribution and the corresponding cumulative distribution
scale as $f_k\approx4/k^3$ and $F_k\approx2/k^2$.
Applying the extreme-value argument to the latter estimate,
we predict that the largest degree grows as
\beq
\K(n)\sim n^{1/2}.
\label{kbasim}
\eeq

In our quest of the distribution of the largest degree $\K(n)$,
our first approximation relies on stationary extreme-value statistics.
We recall that this approach cannot be exact,
for the three reasons exposed in Section~2.1.
Within this approach,
the distribution $\phi_k=\prob{\K(n)=k}$ of the largest degree can be estimated
by means of~(\ref{phiserie}).
The first term ($C=1$) is clearly leading for $k$ large,
where $f_k$ is small.
We thus obtain a distribution of the form $\phi_k\approx nf_k\exp(-nF_k)$, i.e.,
\beq
\phi_k\approx\frac{4n}{k^3}\,\e^{-2n/k^2}.
\label{ksev}
\eeq
In other words, the ratio
\beq
Y(n)=\frac{\K(n)}{n^{1/2}}
\eeq
is predicted to have a non-trivial asymptotic distribution $\rho_Y$.
Our first approximation to the latter distribution,
given by the stationary approach, reads
\beq
\rho_{Y,\st}(Y)\approx\frac{4}{Y^3}\,\e^{-2/Y^2}.
\label{rhosev}
\eeq
This is the Fr\'echet distribution
known in extreme-value statistics~\cite{evs}.
We have in particular $\mean{Y}_\st=(2\pi)^{1/2}\approx2.506628$,
whereas the second moment $\mean{Y^2}_\st$ is divergent.
This feature of the stationary approach is clearly incorrect.
Indeed finite-size effects induce a cutoff,
beyond which the degree distribution falls off very fast.

Our second approximation consists in incorporating these finite-size effects.
This approach is however still inexact,
as it complies with point (2) of the discussion of Section~2.1,
albeit not with point (3),
concerning the dependence of the degree of the node on its index.
Finite-size effects on the degree statistics
have been investigated in detail in~\cite{I}.
The distribution $f_k(n)$ of the degree $k$ at time $n$
and the corresponding cumulative distribution $F_k(n)$ scale as
\beq
f_k(n)\approx\frac{4}{k^3}\,\Phi(y),\quad
F_k(n)\approx\frac{2}{k^2}\,\Psi(y),\quad
y=\frac{k}{n^{1/2}},
\label{fss}
\eeq
with $\Phi(y)=\Psi(y)-(y/2)\Psi'(y)$.
Both scaling functions depend on the initial condition.
The known exact expressions of $\Phi(y)$~\cite[Eq.~(3.47)]{I} yield
\beq
\matrix{
\Phi^\A(y)=\erfc\left(\frad{y}{2}\right)
+\frad{y}{\sqrt{\pi}}\left(1+\frad{y^2}{2}\right)\e^{-y^2/4},\hfill\cr
\Psi^\A(y)=\frad{y}{\sqrt{\pi}}\,\e^{-y^2/4}
+\left(1+\frad{y^2}{2}\right)\erfc\left(\frad{y}{2}\right),\hfill\cr
\Phi^\B(y)=\left(1+\frad{y^2}{4}+\frad{y^4}{8}\right)\e^{-y^2/4},\hfill\cr
\Psi^\B(y)=\left(1+\frad{y^2}{2}\right)\e^{-y^2/4}.\hfill}
\label{fssfun}
\eeq
Our second approximation to the asymptotic distribution of $Y$,
given by the stationary approach including finite-size effects, reads therefore
\beq
\rho_{Y,\fss}(Y)\approx\frac{4\Phi(Y)}{Y^3}\,\e^{-2\Psi(Y)/Y^2}
\label{rhofss}
\eeq
(with `FSS' for `finite-size scaling').
As expected, this expression is better behaved
than the Fr\'echet distribution~(\ref{rhosev}).
It falls off as $\exp(-Y^2/4)$, so that all the moments of~$Y$ are finite.
From a quantitative viewpoint, a numerical evaluation of the appropriate
integrals yields the estimates
$\mean{Y}_\fss^\A\approx1.842146$ and $\mean{Y}_\fss^\B\approx1.966678$.
It is also of interest to get an estimate for the asymptotic value
of the reduced variance
\beq
v(n)=\frac{\var\K(n)}{\mean{\K(n)}^2}=\frac{\var Y(n)}{\mean{Y(n)}^2}.
\eeq
We obtain similarly $v_\fss^\A\approx0.161205$ and $v_\fss^\B\approx0.179513$.

\begin{figure}
\begin{center}
\includegraphics[angle=90,width=.45\linewidth]{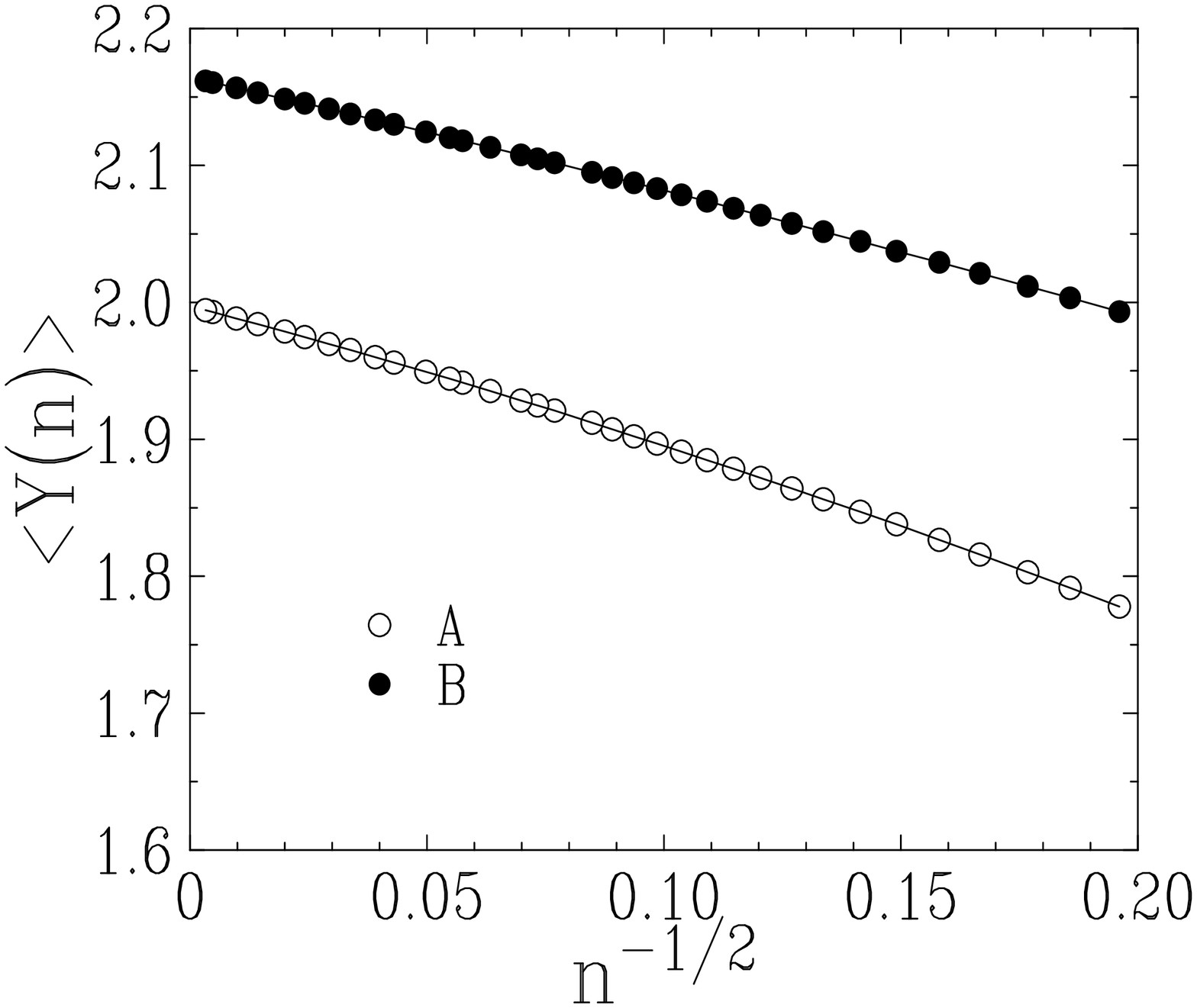}
\includegraphics[angle=90,width=.45\linewidth]{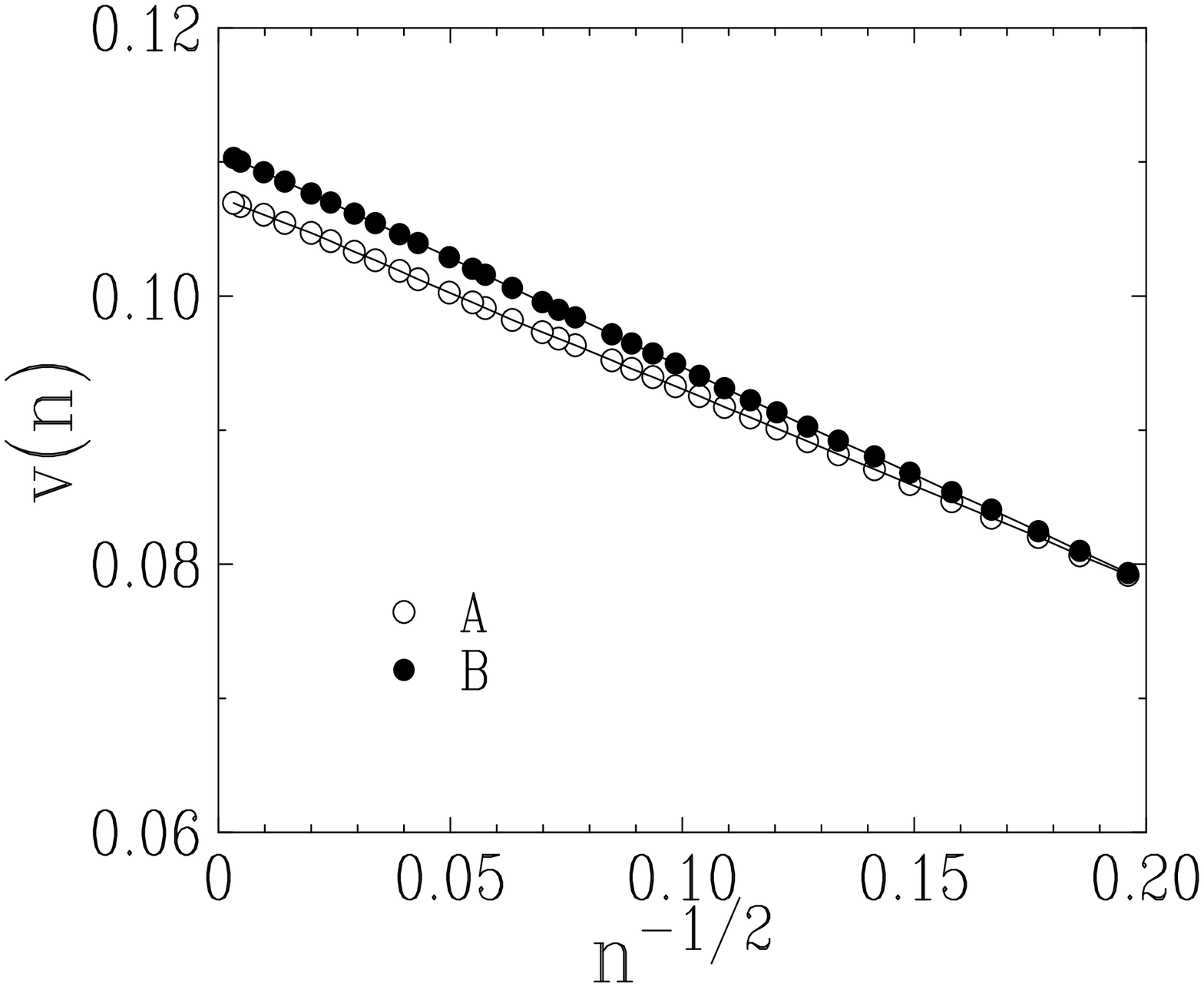}
\caption{\label{figKba}
Statistics of the largest degree in the BA model for both initial conditions.
Left: plot of $\mean{Y(n)}=\mean{\K(n)}/n^{1/2}$ against $n^{-1/2}$.
Right: plot of the reduced variance $v(n)$ against $n^{-1/2}$.
The data have the asymptotic values
$\mean{Y}_\infty^\A\approx2.00$, $\mean{Y}_\infty^\B\approx2.16$,
$v_\infty^\A\approx0.108$ and $v_\infty^\B\approx0.111$.}
\end{center}
\end{figure}

\begin{figure}
\begin{center}
\includegraphics[angle=90,width=.45\linewidth]{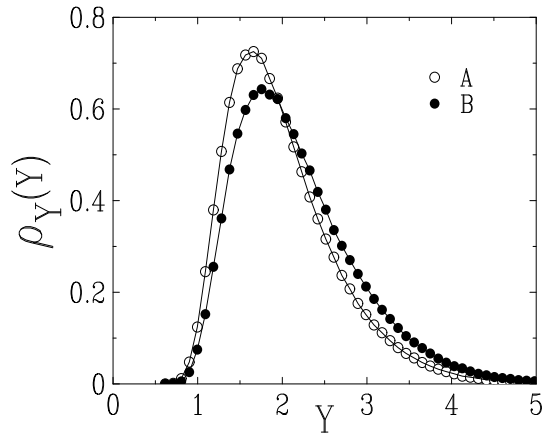}
\caption{\label{figkba}
Plot of the `asymptotic' ($n=10^5$) distribution $\rho_Y$
of the rescaled largest degree $Y$,
for the BA model with both initial conditions.
Data are sorted in bins of width $\delta k=30$.
The difference between both initial conditions is small but significative.}
\end{center}
\end{figure}

Figure~\ref{figKba} shows a plot of the mean rescaled largest degree
$\mean{Y(n)}=\mean{\K(n)}/n^{1/2}$ (left)
and of the corresponding reduced variance $v(n)$ (right),
against $n^{-1/2}$, for both initial conditions.
Throughout this section on the BA model,
simulations are realized by using the redirection algorithm
proposed in~\cite{kr} and recalled in the beginning of Section~4.
Data are averaged over $10^6$ different histories up to time $n=10^5$.
For both quantities shown in the figure,
as well as other ones to be presented below,
the data nicely converge to asymptotic values which can be measured accurately,
with leading corrections proportional to $1/k_\star(n)\sim n^{-1/2}$.
An explanation for this form of corrections will be given in Section~3.3.
The measured asymptotic values read
$\mean{Y}_\infty^\A\approx2.00$, $\mean{Y}_\infty^\B\approx2.16$,
$v_\infty^\A\approx0.108$ and $v_\infty^\B\approx0.111$.
The estimates $\mean{Y}_\fss$ for the mean degree are satisfactory:
they are only some 10\% below the observed values,
and correctly predict that $\mean{Y}_\infty^\B$ is larger
than $\mean{Y}_\infty^\A$.
The approach is however less accurate for the reduced variance,
as the estimates $\mean{v}_\fss$ are too large at least by some 50\%.
Figure~\ref{figkba} shows a histogram plot of the distribution
of the rescaled largest degree $Y(n)$.
The large value of time ($n=10^5$) ensures that the data virtually follow
the asymptotic distributions $\rho_Y^\A$ and $\rho_Y^\B$.
Data have been sorted in bins of width $\delta k=30\sim\mean{\K}/20$,
as $\mean{\K}^\A\approx631$ and $\mean{\K}^\B\approx684$.

\subsection{Number of leads}

The number of leads $\L(n)$ up to time $n$
can again be estimated by considering the effect of co-leaders.
The probability of a lead change is still proportional to $C(n)-1$,
whereas each co-leader now has a probability $\K(n)/(2n)$
to attach the $(n+1)$--st node.

Within the framework of stationary extreme-value statistics,
with $f_k\approx4/k^3$ and $F_k\approx2/k^2$,
the expression~(\ref{phiserie}) shows that the probability of having $C=2$
is small, whereas that of having $C=3$ is still smaller, and so on.
In other words, with respect to the UA model, co-leaders are now more rare,
but the few ones that are present are more efficient.
We thus obtain the estimate
\beq
\mean{\L(n+1)}-\mean{\L(n)}
\approx\sum_{k=1}^\infty\bin{n}{2}f_k^2\,\exp(-nF_k)\,\frac{k}{2n}.
\eeq
For a large time $n$, using the above estimates,
the right-hand side can be shown to boil down to $1/(2n)$.
We thus obtain the same logarithmic growth for the mean number of leads
as for the UA model~\cite{lkr}:
\beq
\mean{\L(n)}\approx A_\L\,\ln n.
\label{lba}
\eeq
Furthermore the stationary approach yields the estimate
$A_{\L,\st}=1/2$ for the amplitude.

\begin{figure}
\begin{center}
\includegraphics[angle=90,width=.45\linewidth]{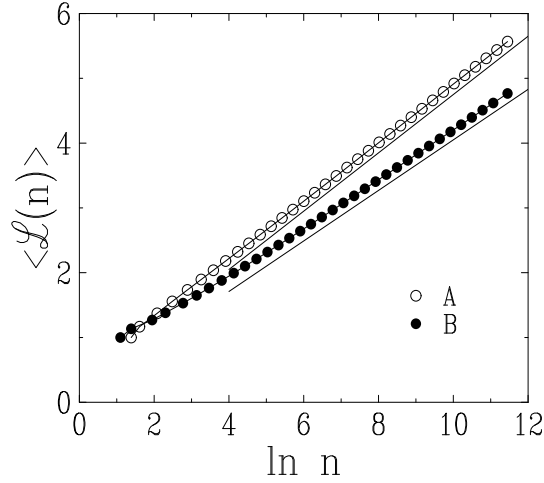}
\caption{\label{figlba}
Plot of the mean number $\mean{\L(n)}$ of leads against $\ln n$,
for the BA model with both initial conditions.
The full lines have slopes $A_\L^\A\approx0.45$ and $A_\L^\B\approx0.39$.}
\end{center}
\end{figure}

Figure~\ref{figlba} shows a plot of $\mean{\L(n)}$ against $\ln n$
for both initial conditions.
The data are observed to follow the logarithmic law~(\ref{lba}),
with two different amplitudes, $A_\L^\A\approx0.45$ and $A_\L^\B\approx0.39$,
in the same range as the estimate $A_{\L,\st}$ of the stationary approach.

\subsection{Index of the leader}

The essential ingredient is again the mean index of a node of given degree $k$.
The numerator $g_k(n)$ of the expression~(\ref{meanid})
can be shown to obey the recursion
\beq
(n+1)Z(n)g_k(n+1)=n(k-1)g_{k-1}(n)+n(Z(n)-k)f_k(n)+(n+1)\delta_{k,1}.
\eeq
In the stationary regime, we have again $g_k(n)\approx n\gamma_k$, with
\beq
(k+4)\gamma_k=(k-1)\gamma_{k-1}+2\delta_{k,1},
\eeq
hence
\beq
\gamma_k=\frac{48}{k(k+1)(k+2)(k+3)(k+4)}
\label{gba}
\eeq
and finally
\beq
\mean{i_k(n)}\approx\frac{12n}{(k+3)(k+4)}.
\label{ikba}
\eeq
Replacing $k$ in this expression by the power law~(\ref{kbasim})
for the typical degree of the leader,
we obtain a result of order unity.
We are thus led to the conclusion that the index of the leader
typically stays finite,
and especially that its mean value has a finite limit $\mean{I}_\infty$.
Within the framework of stationary extreme-value statistics,
evaluating $\mean{I}_\st$ as $\mean{12n/k^2}$ using~(\ref{ksev}),
we obtain the estimate $\mean{I}_\st=6$.
Finally, as the expression~(\ref{ikba}) admits an expansion in powers of $1/k$,
the stationary approach suggests that the leading finite-time corrections
to the limit $\mean{I}_\infty$ are proportional to $1/\K(n)\sim n^{-1/2}$.

\begin{figure}
\begin{center}
\includegraphics[angle=90,width=.45\linewidth]{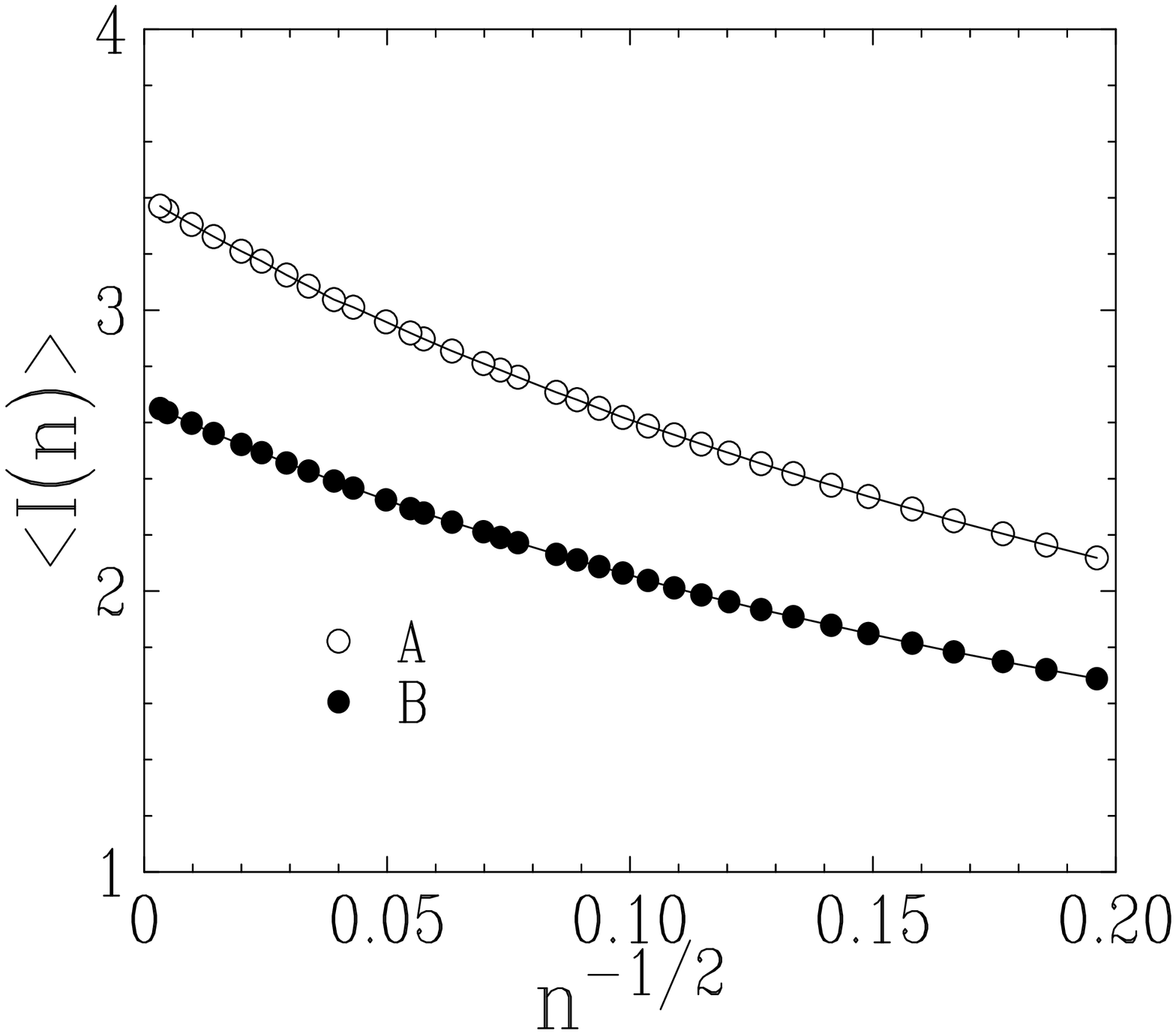}
\caption{\label{figjba}
Plot of the mean index of the leader $\mean{I(n)}$ against $n^{-1/2}$,
for the BA model with both initial conditions.
The asymptotic values read
$\mean{I}_\infty^\A\approx3.40$ and $\mean{I}_\infty^\B\approx2.67$.}
\end{center}
\end{figure}

Figure~\ref{figjba} shows a plot of the mean index of the leader
$\mean{I(n)}$ against $n^{-1/2}$, for both initial conditions.
The data converge to well-defined asymptotic values
$\mean{I}_\infty^\A\approx3.40$ and $\mean{I}_\infty^\B\approx2.67$,
whereas the leading corrections are observed to be linear in~$n^{-1/2}$.

\subsection{Number of distinct leaders and lead persistence probability}

The behavior of these last two quantities can be deduced from
what is already known by means of the following line of reasoning.
The number $D(n)$ of distinct leaders up to time $n$
is at most equal to the largest index of the leader up to time $n$,
$\I(n)={\rm max}(I(1),\dots,I(n))$.
Loosely speaking, $D(n)$ cannot grow typically faster
than the index of the leader.
It is therefore expected to stay of order unity,
and to share with $I(n)$ the property of having a finite asymptotic
mean $\mean{D}_\infty$.
More generally, the whole distribution of the number of distinct leaders
is expected to converge to a non-trivial distribution~$\rho_D$
in the limit of long times.
Finally, as the lead persistence probability $S(n)$
is nothing but the probability that $D(n)$ equals one,
it is also expected to have a finite limit~$S_\infty=\rho_D(1)$.

\begin{figure}
\begin{center}
\includegraphics[angle=90,width=.45\linewidth]{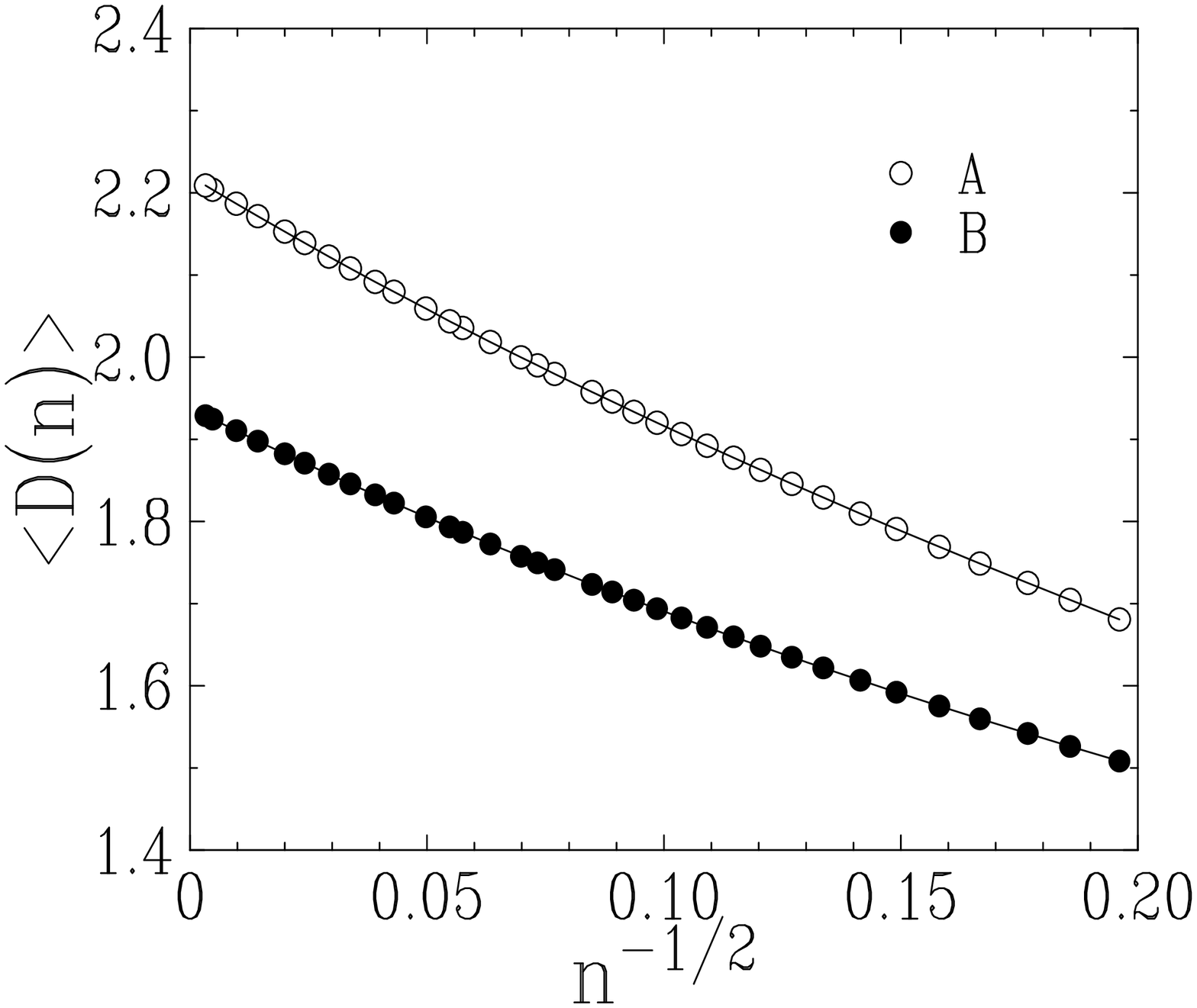}
\includegraphics[angle=90,width=.45\linewidth]{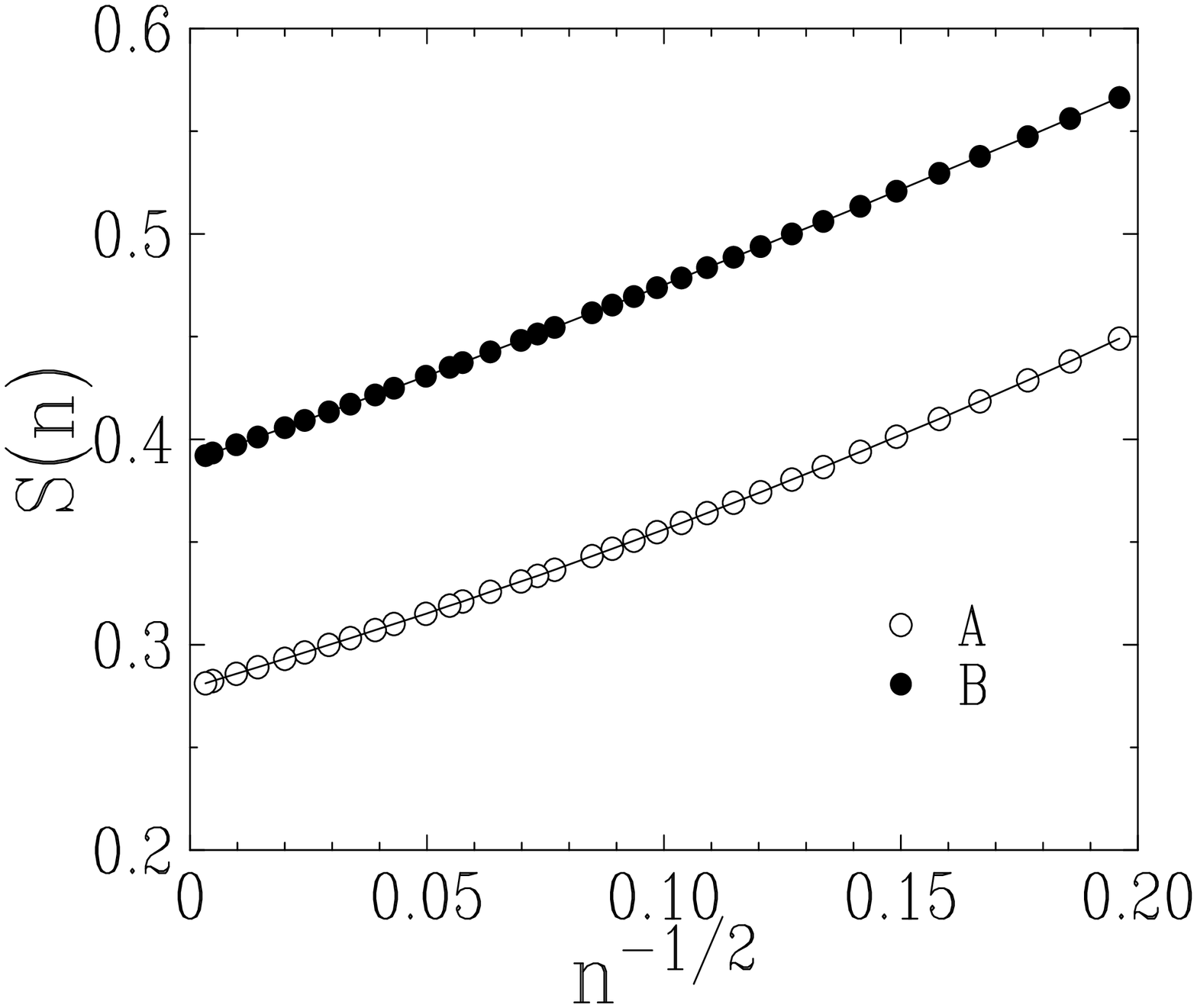}
\caption{\label{figdba}
Left: plot of the mean number $\mean{D(n)}$ of distinct leaders
against $n^{-1/2}$.
Right: plot of the lead persistence probability $S(n)$ against $n^{-1/2}$.
Data for the BA model with both initial conditions have asymptotic values
$\mean{D}_\infty^\A\approx2.22$, $\mean{D}_\infty^\B\approx1.94$,
$S_\infty^\A\approx0.279$ and $S_\infty^\B\approx0.389$.}
\end{center}
\end{figure}

\begin{figure}
\begin{center}
\includegraphics[angle=90,width=.45\linewidth]{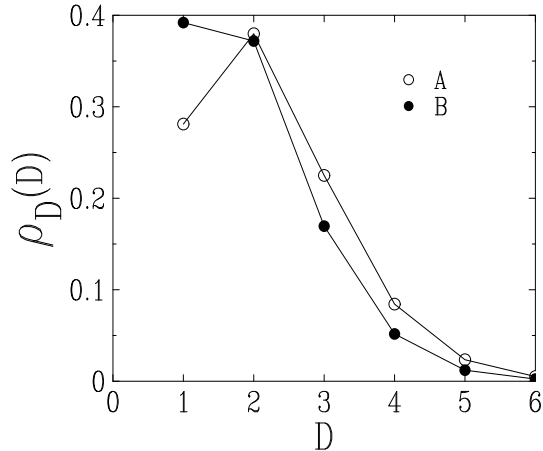}
\caption{\label{figdd}
Plot of the `asymptotic' ($n=10^5$) distribution $\rho_D$
of the number $D$ of distinct leaders,
for the BA model with both initial conditions.}
\end{center}
\end{figure}

Figure~\ref{figdba} shows a plot of the mean number $\mean{D(n)}$
of distinct leaders (left) and of the persistence probability $S(n)$ (right)
against $n^{-1/2}$, for both initial conditions.
The data converge to the asymptotic values
$\mean{D}_\infty^\A\approx2.22$, $\mean{D}_\infty^\B\approx1.94$,
$S_\infty^\A\approx0.279$ and $S_\infty^\B\approx0.389$.
Figure~\ref{figdd} shows a plot of the full distribution
of the number of distinct leaders for $n=10^5$
and both initial conditions.
The plotted data are virtually equal to the asymptotic distribution $\rho_D$.

\section{The general preferential attachment (GPA) model}

In the general preferential attachment (GPA) model,
the probability of attachment to a node
is proportional to the sum of the degree of that node
and of a constant parameter $c>-1$,
representing the initial attractiveness of a node~\cite{dms1}.
This parameter is relevant, as it yields the continuously varying exponents
\beq
\gamma=c+3,\quad\nu=\frac{1}{c+2}.
\label{apgex}
\eeq
The BA and UA models are respectively recovered when $c=0$ and $c\to\infty$.

More precisely, in the GPA model the attachment probability of node $n$
to node $i$ reads
\beq
p_{n,i}=\frac{k_i(n-1)+c}{Z(n-1)}.
\label{pgpa}
\eeq
The partition function in the denominator reads $Z(n)=2L(n)+cn$.
It therefore depends on the initial condition according to
\beq
Z^\A(n)=(c+2)n-2,\quad Z^\B(n)=(c+2)n-1.
\label{zgpa}
\eeq

The stationary degree distribution of the GPA model
can be derived along the lines of Section~2.1.
The distribution $f_k(n,i)$ of the degree $k_i(n)$ of node $i$ at time $n$
obeys the recursion
\beq
Z(n)f_k(n+1,i)=(k+c-1)f_{k-1}(n,i)+(Z(n)-k-c)f_k(n,i),
\eeq
with $f_k(i,i)=\delta_{k,1}$ for $i\ge2$.
The distribution of the degree of an arbitrary node
therefore obeys the recursion
\beq
(n+1)Z(n)f_k(n+1)=n(k+c-1)f_{k-1}(n)+n(Z(n)-k-c)f_k(n)+\delta_{k,1}.
\eeq
Finally, the stationary degree distribution $f_k$ obeys the recursion
\beq
(k+2c+2)f_k=(k+c-1)f_{k-1}+(c+2)\delta_{k,1},
\eeq
hence~\cite{dms1,krl}
\beq
f_k=\frac{(c+2)\Gamma(2c+3)\Gamma(k+c)}{\Gamma(c+1)\Gamma(k+2c+3)}.
\label{gfstat}
\eeq
This expression has a power-law decay at large $k$:
\beq
f_k\approx\frac{(c+2)\Gamma(2c+3)}{\Gamma(c+1)}\,k^{-(c+3)}.
\label{gfkas}
\eeq
In the regime where $k$ and $c$ are both large and comparable,
the expression~(\ref{gfstat}) assumes the scaling form~\cite[Eq.~(4.29)]{I}:
\beq
f_k\sim\exp(-c\,\phi(\kappa)),
\label{fssau}
\eeq
with $\kappa=k/c$ and
\beq
\phi(\kappa)=(\kappa+2)\ln(\kappa+2)-(\kappa+1)\ln(\kappa+1)-2\ln 2.
\eeq
The linear behavior $\phi(\kappa)\approx\kappa\ln 2$ as $\kappa\to0$
matches the exponential fall-off~(\ref{fua}) of the UA model,
whereas the logarithmic growth $\phi(\kappa)\approx\ln\kappa+1-2\ln 2$
as $\kappa\to\infty$ matches the power-law decay~(\ref{gfkas}).

The GPA model can be simulated very efficiently (in time $n$)
by means of the following {\it redirection algorithm}~\cite{kr}.
At time $n$, in order to attach node $n$,
an earlier node $i=1,\dots,n-1$ is chosen uniformly,
and node $n$ is attached either to node $i$ itself with probability $1-\nu$,
or to the {\it ancestor} $a(i)$ of node $i$ with probability $\nu$.
The ancestor of $i$ is the node $a(i)=1,\dots,i-1$ to which
node $i$ has attached at time $i$.
The attachment probabilities~(\ref{pgpa}) are recovered if the redirection
probability $\nu$ is chosen to be equal to the growth exponent~(\ref{apgex}),
hence the notation.
In particular, the linear attachment probabilities of the BA model
are recovered when the redirection probability is $\nu=1/2$.
Initial conditions are implemented by requiring that some {\it primordial nodes}
have no ancestor~\cite{kr}.
For Case~A, nodes 1 and 2 have no ancestor,
whereas node~1 is the ancestor of node~3.
For Case~B, node~1 has no ancestor, whereas it is the ancestor of node~2.
The numbers of primordial nodes can be read off
from the expressions~(\ref{zgpa}) of the partition function:
these are the integers which appear after the minus signs.

For any finite $c>-1$, i.e., any value $0<\nu<1$ of the growth exponent,
the GPA model behaves in every respect
similarly to the BA model (corresponding to $\nu=1/2$),
investigated in Section~3.
The largest degree $\K(n)$ grows as $n^\nu$.
Setting $\K(n)=Y(n)\,n^\nu$,
the rescaled variable $Y(n)$ has a limiting distribution $\rho_Y$,
which depends on $\nu$ and on the initial condition.
The mean $\mean{Y(n)}$ and the reduced variance $v(n)$
converge to finite limits $\mean{Y}_\infty$ and $v_\infty$,
with leading corrections proportional to $1/k_\star(n)\sim n^{-\nu}$.
The mean number of leads scales as $\mean{\L(n)}\approx A_\L\ln n$,
where the amplitude $A_\L$ depends on $\nu$ and on the initial condition.
Finally, the mean index of the leader, the mean number of distinct leaders,
and the lead persistence probability
have finite limiting values $\mean{I}_\infty$, $\mean{D}_\infty$ and $S_\infty$.
The leading corrections to these limits are again in $n^{-\nu}$.
The convergence is therefore very slow at small~$\nu$,
where a crossover to the UA model is observed (see below).
In practice limiting values cannot be extrapolated
from data for reasonable times ($n=10^5$ -- $10^6$)
with enough accuracy for $\nu<0.1$.

\begin{figure}
\begin{center}
\includegraphics[angle=90,width=.45\linewidth]{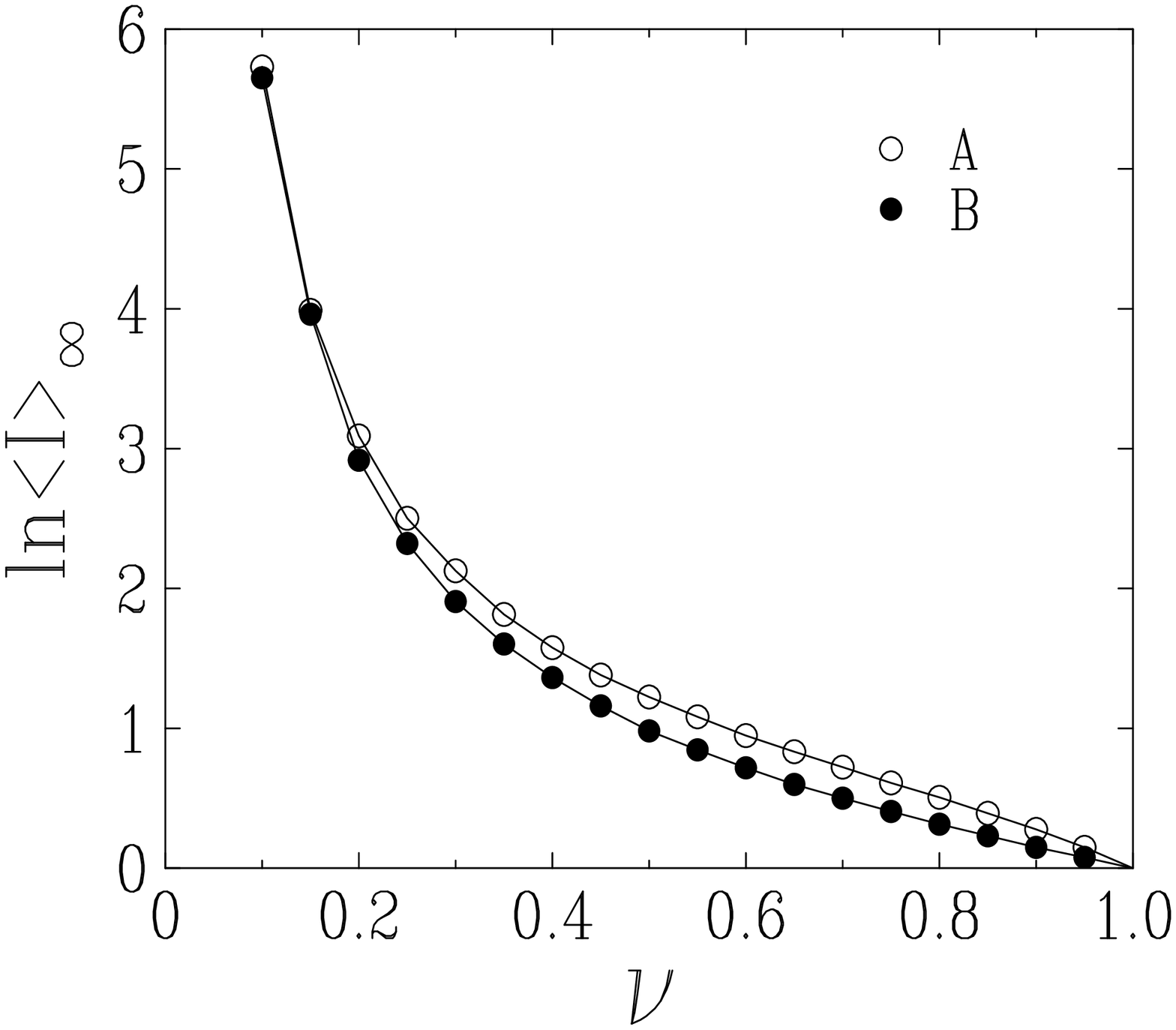}
\includegraphics[angle=90,width=.45\linewidth]{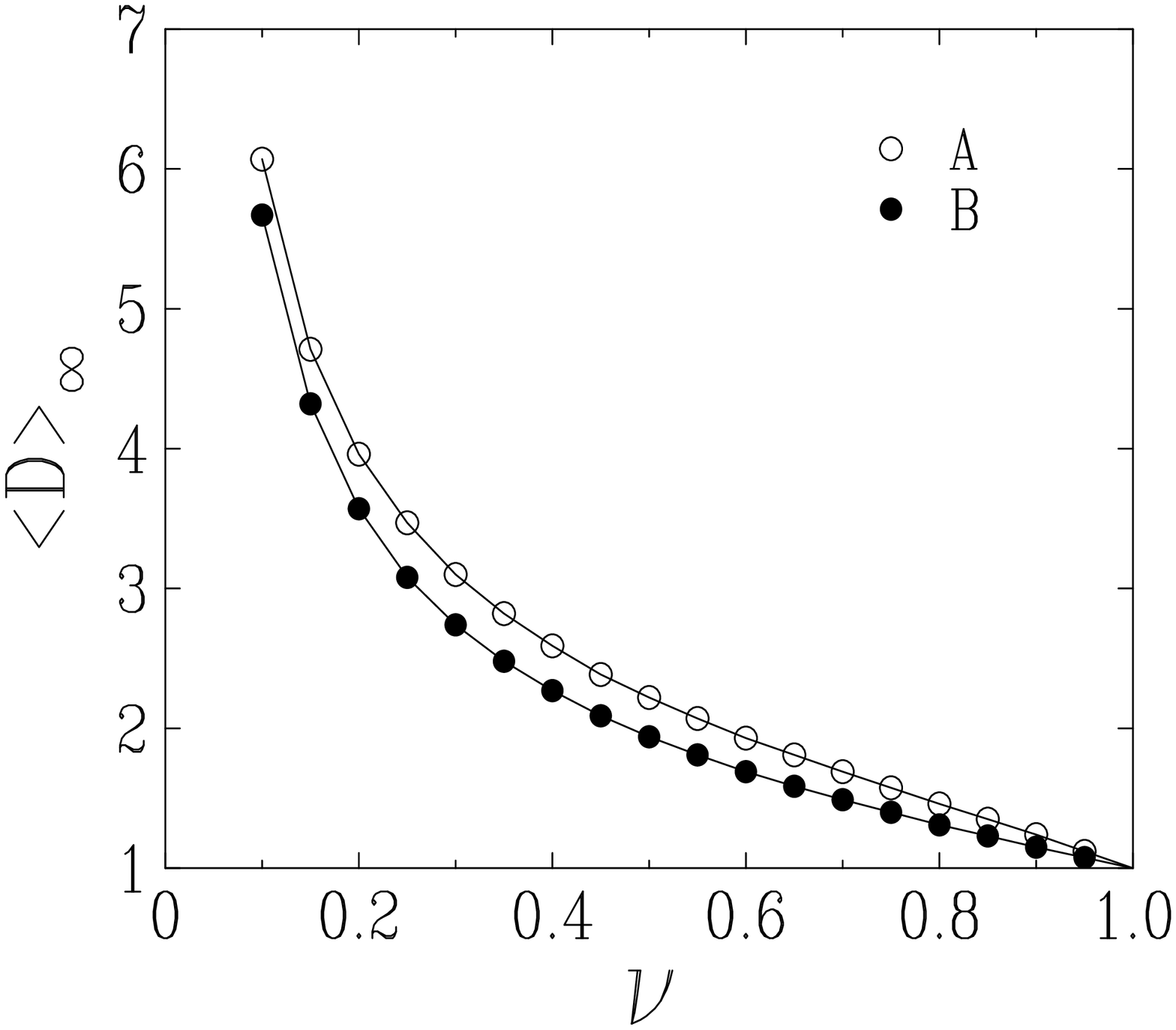}

\includegraphics[angle=90,width=.45\linewidth]{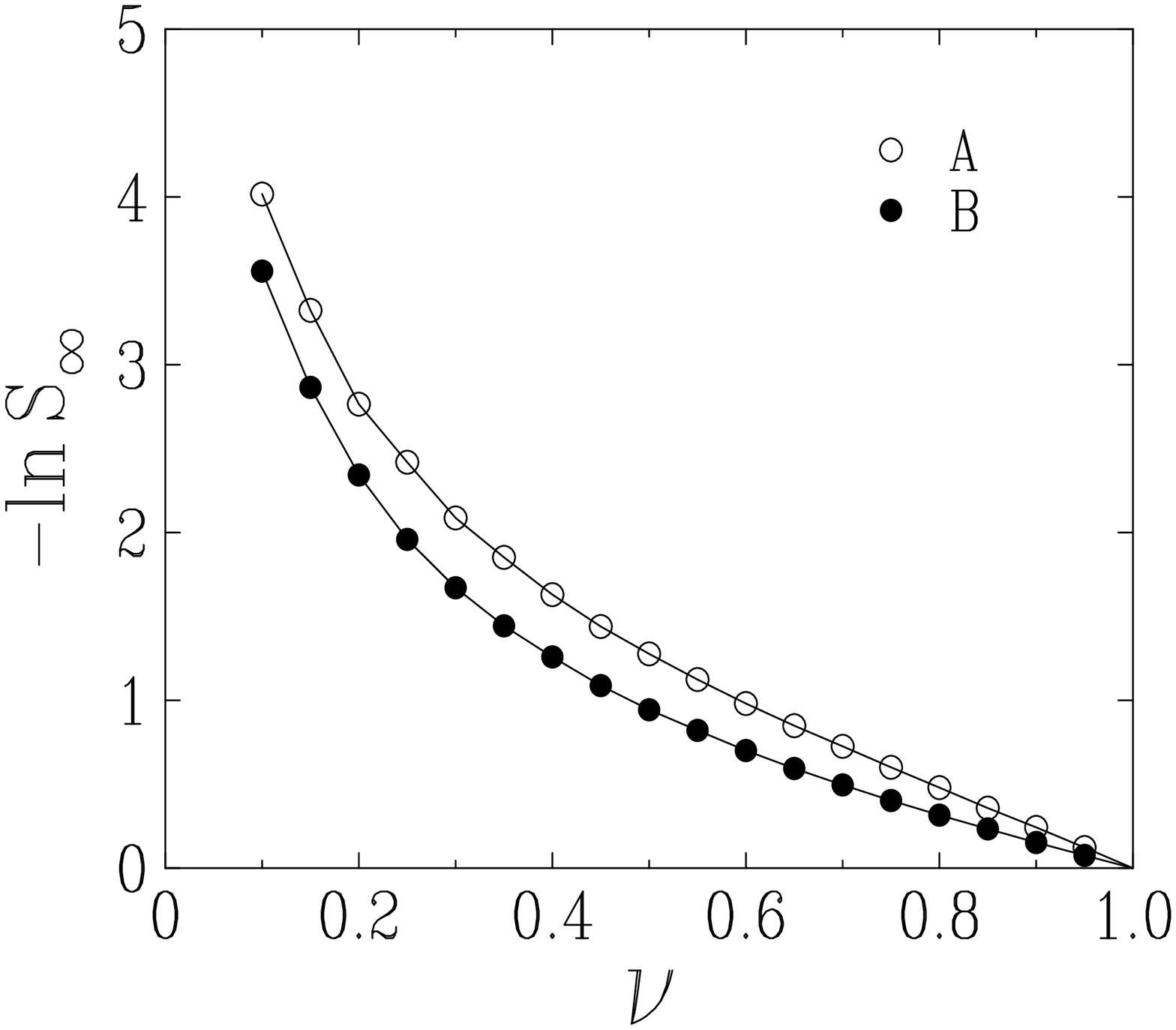}
\includegraphics[angle=90,width=.45\linewidth]{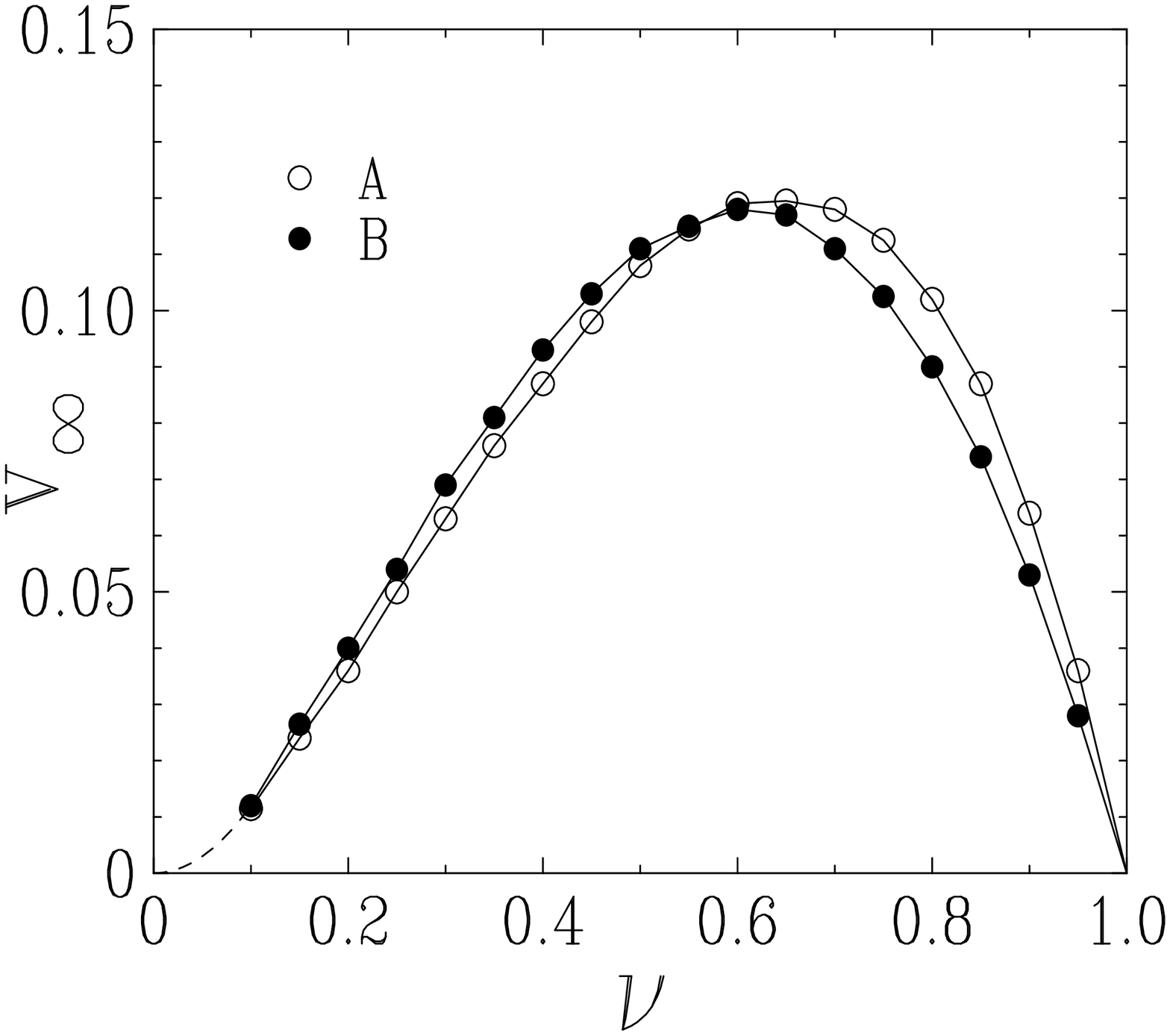}
\caption{\label{figapg}
Plot of the four key asymptotic quantities of the GPA model
with both initial conditions:
$\ln\mean{I}_\infty$, $\mean{D}_\infty$, $-\ln S_\infty$ and $v_\infty$,
against the growth exponent $\nu=1/(c+2)$.}
\end{center}
\end{figure}

The dependence of the key asymptotic quantities
$\mean{I}_\infty$, $\mean{D}_\infty$, $S_\infty$ and $v_\infty$
on the growth exponent $\nu$
for both initial conditions is shown in Figure~\ref{figapg}.
The data corresponding to the BA model, discussed in Section~3,
are recovered for $\nu=1/2$, i.e., right in the middle of the plots.
The behavior at both endpoints also deserves some attention.

\noindent $\bullet$ $\nu\to1$, i.e., $c\to-1$.
In this limit, the GPA model boils down to a ballistic growth model.
Indeed the attachment probability $k+c$ vanishes for $k=1$,
i.e., the initial degree of each node.
As a consequence, with our conventions on the initial states,
the first node attracts all the subsequent ones.
Its degree grows ballistically as $k_1(n)\approx n$,
whereas its lead keeps undisputed.
This picture explains the trivial limiting values of the key quantities
($\mean{I}_\infty\to1$, $\mean{D}_\infty\to1$, $S_\infty\to1$, $v_\infty\to0$).
The leading corrections to these limits are observed
to be proportional to $c+1$, i.e., to $1-\nu$.

\noindent $\bullet$ $\nu\to0$, i.e., $c\to\infty$.
This regime is more subtle than the previous one.
Consider a very large value of the parameter $c\approx1/\nu$.
The scaling behavior~(\ref{fssau}) expresses
that the static degree distribution first falls off exponentially,
as in the UA model (for $k\ll c$), and then crosses over (for $k\sim c$)
to a power-law fall-off with a large exponent (for $k\gg c$).
Assume now that the degree $k$ under consideration
grows as the largest degree $\K(n)\sim\ln n$ in the UA model (see~(\ref{kln})).
The above crossover then takes place at an exponentially large time~$n$,
such that $\ln n$ becomes comparable to $c$.
As a consequence, a rough estimate of the behavior of the key quantities
as $\nu\to0$ can be obtained by replacing $\ln n$ by $1/\nu$
in the large-time behavior of the same quantities for the UA model.
We have seen in Section~2
that $\mean{I(n)}$ grows as a power law (see~(\ref{jua})),
whereas $S(n)$ falls off as a power law (see~(\ref{suares})),
and $\mean{D(n)}$ grows logarithmically (see~(\ref{meandua})).
Finally, $\mean{\K(n)}$ grows logarithmically,
whereas $\var\K(n)$ stays finite, hence $v(n)\sim1/(\ln n)^2$.
We thus obtain the estimates
\beq
\ln\mean{I}_\infty\sim\mean{D}_\infty\sim-\ln S_\infty\sim\frac{1}{\nu},
\quad v_\infty\sim\nu^2.
\eeq
up to multiplicative constants which are not predicted by the
crossover argument.
In spite of their crudeness, the above estimates are in qualitative agreement
with the data shown in Figure~\ref{figapg}.
The first three quantities indeed exhibit a similar kind of divergence
at small~$\nu$, reaching values of order 4 -- 6 for $\nu=0.1$.
The data for the reduced degree variance $v_\infty$
are observed to vanish faster than linearly as $\nu\to0$,
and to be compatible with a quadratic law (dashed line).
Finally, since the reduced variance $v_\infty$ vanishes at both endpoints,
it has a non-trivial maximum
($v_{\infty,\max}^\A\approx0.120$ and $v_{\infty,\max}^\B\approx0.118$).

\section{Discussion}

We have presented a comprehensive study
of the statistics of leaders and lead changes
in growing networks with preferential attachment.
The quantities investigated in this work
concern either the leader at a given time
(degree and identity of the leader, number of co-leaders)
or the whole history of lead changes
(number of lead changes, of distinct leaders, lead persistence probability).
So far very few works had been devoted to the subject
of leaders in this sense,
either in growing networks~\cite{lkr,lm}
or in other random structures~\cite{el,lbk}.
The present paper is also to some extent a continuation
of our previous work~\cite{I} on finite-size effects in the degree statistics.
Both articles are meant to be systematic and consistent with each other.
In particular, the same three different attachment rules (UA, BA, GPA)
and the same two different initial conditions (Cases~A and~B)
are dealt with in parallel in both works.

The first and most natural quantity of this area
is the largest degree $\K(n)$, i.e., the degree of the leader at time $n$.
As recalled in the Introduction,
there are (at least) three ways of defining a time-dependent
characteristic degree scale in a scalefree network (see e.g.~\cite{dms1,bpv,I}):

\noindent (i) the cutoff scale $k_\star(n)$,
beyond which the stationary degree distribution $f_k$
is strongly affected by finite-size effects;

\noindent (ii) the largest degree $\K(n)$;

\noindent (iii) the degree $k_1(n)$ of a node
with any fixed index, say the first one~($i=1$) for definiteness.

These three degree scales have the same growth law as a function of time $n$,
i.e., a logarithmic growth for the UA model,
and a power law with exponent~$\nu$ in the scalefree case.
The agreement between definitions (ii) and (iii) is rather natural,
as the first node has the highest chance of being the leader.
The agreement between definitions (i) and (ii), although less obvious,
can be explained within the framework
of what we have called `stationary extreme-value statistics'.
Although this approach cannot be exact (see Section~2.1),
it however turns out to correctly predict the order of magnitude of $\K(n)$
by the condition that the stationary probability for the degree~$k$
to be larger than $\K(n)$, i.e., the cumulative probability $F_k$,
be of order $1/n$.
This criterion implies that $\K(n)$ is comparable to $k_\star(n)$,
where $F_k$ drops more or less suddenly from large to small values.
The above heuristic picture can be complemented with quantitative results.
In the UA model, the three degree scales grow logarithmically with time $n$,
albeit with different (exactly known) prefactors,
and different scalings for fluctuations.
The degree of the first node is asymptotically distributed
according to a Poissonian law~\cite{lkr,I,kr}
with $\mean{k_1(n)}\approx\ln n+\euler$.
The largest degree grows as $\K(n)\approx\ln n/\ln 2$~(see~(\ref{kln})),
whereas fluctuations around this typical value are of order unity.
They are indeed given by a `discrete Gumbel law', studied in Appendix~A.
Finally, the finite-time cutoff in the degree distribution takes place around
$k_\star(n)\approx 2\ln n$,
in a narrow range whose width scales as $(\ln n)^{1/2}$~\cite{I}.
In the BA model, the above degree scales grow as $n^{1/2}$,
whereas fluctuations remain broad, i.e., of order unity in relative value.
The degree of the first node is asymptotically distributed
according to a geometric law~\cite{lkr,I,kr}
with $\mean{k_1(n)}\approx n^{1/2}$.
The largest degree scales as $\K(n)\approx Y\,n^{1/2}$,
where the rescaled variable $Y$ has a non-trivial
limiting distribution $\rho_Y$,
which depends on the initial condition~(see Figure~\ref{figkba}).
This distribution is qualitatively different from the Fr\'echet distribution
predicted by `stationary extreme-value statistics',
whereas the Gumbel distribution put forward in~\cite{lm}
cannot be more than a good fit of over some range.
Finally, the cutoff scale $k_\star(n)$
is also smeared into a non-trivial finite-size scaling law,
which also depends on the initial condition~(see~(\ref{fss})).
The same kind of asymptotic results holds more generally for the GPA model,
where the growth exponent $1/2$ of the BA model
is replaced by the continuously varying exponent $\nu=1/(c+2)$.

Our results on the other observables,
especially those concerning the whole history of a growing network
(number of lead changes, of distinct leaders, lead persistence probability),
are the outcome of a blend of heuristic reasoning and accurate numerical work.
In the scalefree case (BA and GPA models),
the present study confirms and makes more quantitative
the picture emerging from the pioneering work
of Krapivsky and Redner~\cite{lkr}.
The finiteness of $\mean{I}_\infty$ and $\mean{D}_\infty$
means that a typical infinite history of a scalefree network
involves finitely many distinct leaders,
which are chosen among the oldest nodes.
For the BA model,
the numerical results of Section~3, recalled in Table~\ref{batable},
show that the number of distinct leaders is in fact quite small
(around two on average),
whereas the index of the leader is hardly larger (around three on average).
The entire history even consists of a single lead with an appreciable
lead persistence probability~$S_\infty$ (around one third).
As already underlined in~\cite{lkr},
this is a strong form of the {\it rich-get-richer} principle.

\begin{table}
\begin{center}
\begin{tabular}{|c|c|c|c|c|c|}
\hline
Quantity&$\mean{Y}_\infty$&$v_\infty$&$\mean{I}_\infty$
&$\mean{D}_\infty$&$S_\infty$\\
\hline
Case~A&2.00&0.108&3.40&2.22&0.279\\
\hline
Case~B&2.16&0.111&2.67&1.94&0.389\\
\hline
$\Delta(\%)$&7&3&21&13&28\\
\hline
\end{tabular}
\end{center}
\caption{Dependence on the initial condition (Case~A or Case~B)
of the asymptotic large-time values of quantities
of interest for the BA model, obtained by an accurate extrapolation
of numerical data (see Section~3):
mean reduced largest degree $\mean{Y}_\infty$,
reduced variance $v_\infty$ of the largest degree,
mean index $\mean{I}_\infty$ of the leader,
mean number $\mean{D}_\infty$ of distinct leaders,
and lead persistence probability $S_\infty$.
The last row gives the relative difference between data
for both initial conditions,
expressed as a percentage of the larger figure.}
\label{batable}
\end{table}

The lead persistence probability and other key quantities
exhibit a weak but significative dependence on the initial condition,
as demonstrated by the data of Table~\ref{batable}.
The feature that scalefree networks remember their whole past forever,
and especially their infancy,
already visible in finite-size effects on the degree statistics,
is thus fully confirmed as a general phenomenon.

Finally, the study of the statistics of leaders and lead changes
can be expected to shed some new light onto more complex models.
We think especially of the Bianconi-Barab\'asi (BB) model~\cite{bb},
an extension of the BA model where the attachment rule
involves both dynamical variables (the node degrees $k_i(n)$)
and quenched disordered ones (the node fitnesses $\eta_i$).
The BB model has the remarkable feature that it may
exhibit a continuous transition,
somewhat analogous to the Bose-Einstein condensation,
or to the condensation transitions which take place
in classical stochastic models such as the zero-range process (ZRP)~\cite{zrp}.
It is convenient to parametrize the node fitnesses as {\it activated variables:}
$\eta_i=\exp(-\varepsilon_i/T)$,
where the activation energies $\varepsilon_i$
are drawn from a temperature-independent distribution.
The BA model is recovered in the infinite-temperature limit,
whereas the opposite zero-temperature limit yields an interesting
{\it record-driven growth process}, investigated at length in~\cite{usrecords}.
Depending on the distribution of the $\varepsilon_i$,
the BB model may have a low-temperature condensed phase
below some finite condensation temperature $T_c$.
The dynamics of the model in its condensed phase,
already tackled in~\cite{fb}, is not fully understood yet.
The statistics of leaders seems to be an appropriate tool,
as the condensate should naturally appear as a long lived leader.
We intend to return to this problem in a near future.

\appendix

\section{Extreme-value statistics for geometric integer variables}

This Appendix gives a self-contained exposition of various results concerning
extreme-value statistics for integer variables.
The theory of extreme-value statistics is most currently presented
in the case of real variables with a continuous distribution~\cite{evs}.
The peculiarities of discrete (e.g.~integer) distributions
have been underlined in~\cite{anderson}.

After some general formalism,
we study more extensively the case of a large collection
of $n$ independent integer variables $k_i$ ($i=1,\dots,n$),
referred to as degrees, drawn from the common geometric distribution
\beq
f_k=(1-a)a^{k-1}\quad(k\ge1).
\label{fdef}
\eeq
The parameter $a$ can assume any value in the range $0<a<1$.
The stationary degree distribution~(\ref{fua}) of the UA model
is recovered for $a=1/2$.
We consider successively the largest degree,
$\K(n)={\rm max}(k_1,\dots,k_n)$, and the number $C(n)$ of co-leaders,
i.e., the number of node indices $i$ such that $k_i=\K(n)$.

\subsection*{A1.~Distribution of the largest degree}

Let us start with some general formalism.
The derivation of the distribution of the largest degree $\K(n)$
follows the usual line of reasoning of extreme-value statistics~\cite{evs}.
Introducing the cumulative distribution
\beq
F_k=\prob{k_i\ge k}=\sum_{j=k}^\infty f_j,
\label{Fdef}
\eeq
so that
\beq
f_k=F_k-F_{k+1},
\label{fdiff}
\eeq
we have
\beq
\prob{\K(n)<k}=(1-F_k)^n.
\label{Fn}
\eeq
The distribution of the largest degree, $\phi_k=\prob{\K(n)=k}$, thus reads
\beq
\phi_k=(1-F_{k+1})^n-(1-F_k)^n.
\eeq
Using~(\ref{fdiff}), this expression can be expanded as
\beq
\phi_k=\sum_{C=1}^n\bin{n}{C}f_k^C(1-F_k)^{n-C}.
\label{phiserie}
\eeq
The integer $C\ge1$ is to be interpreted
as the number of co-leaders~(see~(\ref{joint})).

In the present situation of interest,
i.e., the geometric distribution~(\ref{fdef}),
the above general formulas read
\beq
F_k=a^{k-1}
\label{Fgeo}
\eeq
and
\beq
\phi_k=(1-a^k)^n-(1-a^{k-1})^n.
\label{phiexact}
\eeq

Hereafter we are mostly interested in large values of $n$.
In this regime, the latter expression can be `exponentiated' as follows:
\beq
\phi_k\approx\e^{-na^k}-\e^{-na^{k-1}}.
\label{phiasy}
\eeq
The mean value of $\K(n)$ can thus be recast as
\beq
\mean{\K(n)}\approx\sum_{k=1}^\infty k(\e^{-na^k}-\e^{-na^{k-1}}).
\label{meanK1}
\eeq
The sum is dominated by the values of $k$ such that $na^k$ is of order unity.
We thus obtain the estimate
\beq
\mean{\K(n)}\approx\frac{\ln n}{\abs{\ln a}}.
\label{kest}
\eeq
A more refined analysis goes as follows.
Consider the generating function
\beq
G_n(s)=\mean{\e^{s\K(n)}}
\approx\sum_{k=1}^\infty\e^{sk}(\e^{-na^k}-\e^{-na^{k-1}}).
\label{gsum}
\eeq
The estimate~(\ref{kest}) suggests to set
\beq
n=a^{-j-\xi},\;\;\mbox{i.e.,}\quad\frac{\ln n}{\abs{\ln a}}=j+\xi,
\label{jdef}
\eeq
with $j$ integer and $0\le\xi<1$ for definiteness, and
\beq
k=j+m.
\label{elldef}
\eeq
With these notations, the expression~(\ref{gsum}) can be recast as
\beq
G_n(s)\approx A(s,\xi)\,n^{s/\abs{\ln a}},
\label{gres}
\eeq
with
\beq
A(s,\xi)=\sum_{m=-\infty}^\infty
\e^{s(m-\xi)}(\e^{-a^{m-\xi}}-\e^{-a^{m-1-\xi}}).
\label{adef}
\eeq
The amplitude $A(s,\xi)$ is a periodic function of $\xi$, with unit period.
Its Fourier series can be derived by means of the Poisson summation formula,
in the form
\beq
\sum_{m=-\infty}^\infty f(m-\xi)=\sum_{n=-\infty}^\infty\e^{2\pi\i n\xi}
\int_{-\infty}^\infty f(x)\,\e^{2\pi\i nx}\,\d x.
\eeq
We thus obtain
\beq
A(s,\xi)=\sum_{n=-\infty}^\infty\alpha_n(s)\,\e^{2\pi\i n\xi},
\eeq
with
\beq
\alpha_n(s)=\frac{\e^s-1}{\ln a}\,\Gamma\!\left(\frac{s+2\pi\i n}{\ln a}\right).
\label{alphares}
\eeq
The oscillations in the amplitude $A(s,\xi)$,
described by the Fourier modes with $n\ne0$,
are extremely small, except if $a$ is very close to zero.
The Fourier coefficients indeed fall off exponentially
as $\abs{\alpha_n(s)}\sim\exp(-\pi^2\abs{n}/\abs{\ln a})$, irrespective of $s$.
For $a=1/2$, our case of interest,
this estimate reads $\abs{\alpha_n(s)}\sim(6.549\times 10^{-7})^{\abs{n}}$,
so that even the first amplitudes ($n=\pm1$) are very small.

Asymptotic expressions for the mean and the variance
of the largest degree can be obtained by expanding~(\ref{gres})
to second order near $s=0$.
Neglecting the tiny periodic oscillations in $\xi$,
i.e., replacing the full function $A(s,\xi)$
by the constant amplitude~$\alpha_0(s)$, we are left with
\beq
\mean{\K(n)}\approx\frac{\ln n+\euler}{\abs{\ln a}}+\frac{1}{2},\quad
\var\K(n)\approx\frac{\pi^2}{6(\ln a)^2}+\frac{1}{12},
\label{mvK}
\eeq
where $\euler\approx0.577215$ denotes Euler's constant.

It is worth comparing the present problem to its continuous analogue,
namely the distribution of $\X(n)$,
the largest of a large number~$n$ of i.i.d.~random variables $x_i>0$,
with exponential density $f(x)=\lambda\,\e^{-\lambda x}$,
with the identification $\lambda=\abs{\ln a}$.
This is a standard problem of extreme-value statistics~\cite{evs}.
Setting
\beq
\X(n)=\frac{\ln n+\xi}{\lambda},
\eeq
the reduced variable $\xi$ obeys a Gumbel law with density
\beq
\rho_\xi(\xi)=\e^{-\xi-\e^{-\xi}}.
\eeq
We have $\mean{\e^{s\xi}}=\Gamma(1-s)$,
so that $\mean{\xi}=\euler$ and $\var\xi=\pi^2/6$,
hence
\beq
\mean{\X(n)}\approx\frac{\ln n+\euler}{\lambda},\quad
\var\X(n)\approx\frac{\pi^2}{6\lambda^2}.
\eeq
These results coincide with the first terms of the expressions~(\ref{mvK}),
whereas the second terms there, involving rational numbers,
are intrinsic effects of discreteness.

We will refer to the distribution~(\ref{phiasy}) as a {\it discrete Gumbel law}.
Let us now turn to other features of this distribution.
Using the definitions~(\ref{jdef}),~(\ref{elldef}),
the expression~(\ref{phiasy}) can be recast as
\beq
\phi_k\approx\e^{-a^{m-\xi}}-\e^{-a^{m-1-\xi}}.
\label{phiell}
\eeq
The full distribution exhibits periodic oscillations in $\xi$, with unit period.
These oscillations have been shown to be very small
in the case of global characteristics, such as the mean and variance.
They are however more visible in the shape of the distribution near its maximum.
This shape indeed oscillates between:

\noindent (i) a symmetric triangular form
(\includegraphics[angle=90,height=13pt]{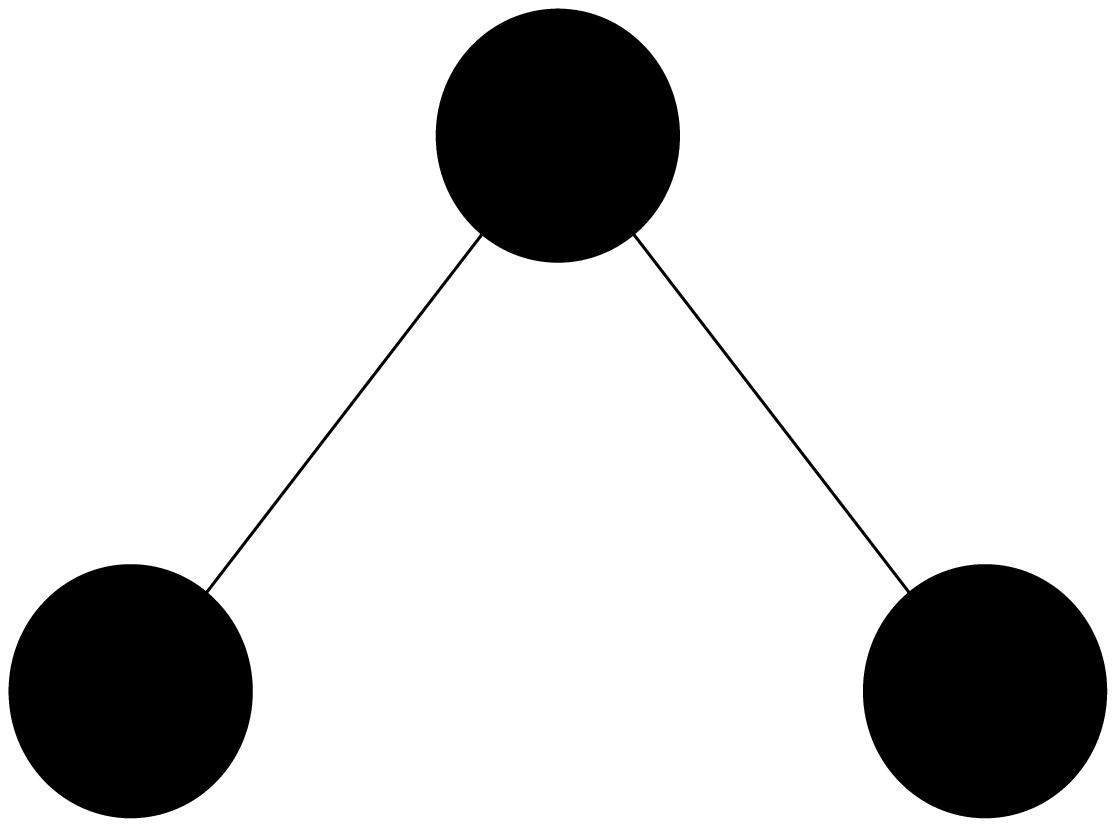})
with a well identified most probable integer,

\noindent (ii) a flat tabular form
(\includegraphics[angle=90,height=13pt]{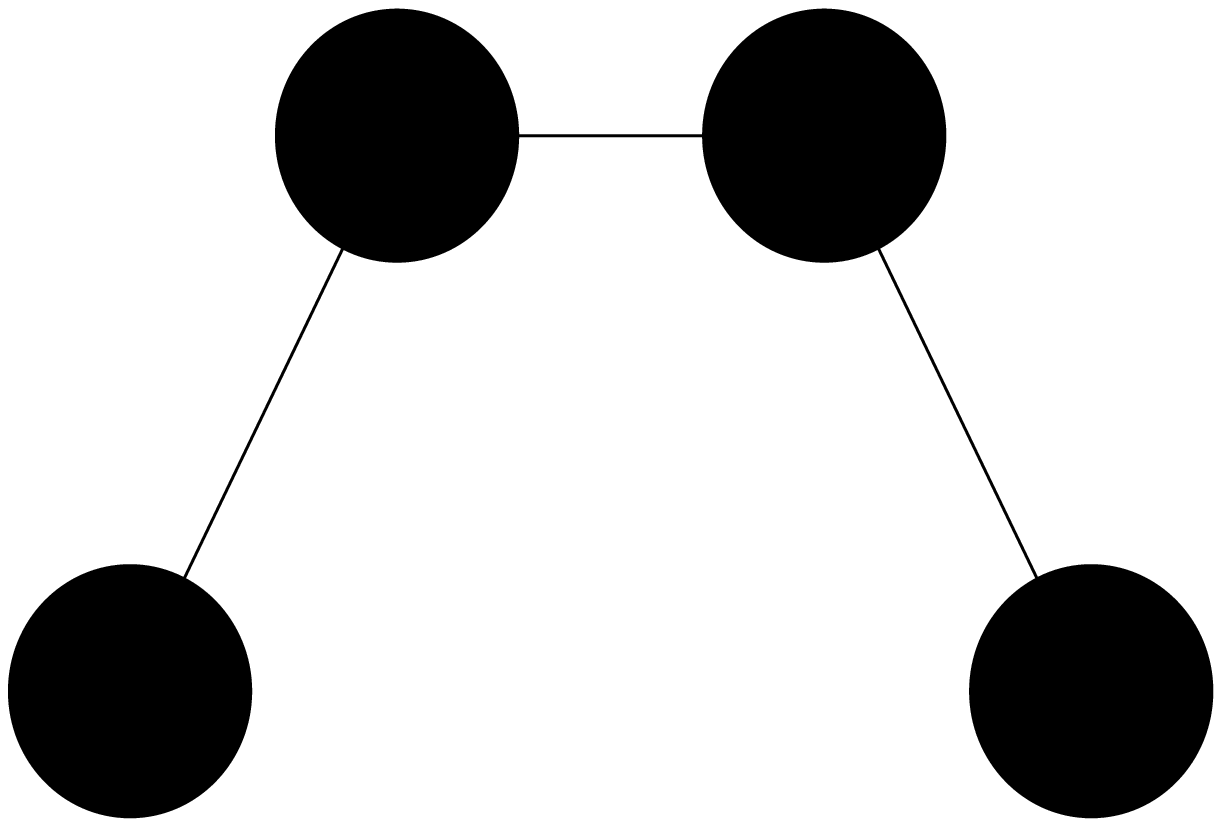})
with two equally probable integers.

\begin{figure}
\begin{center}
\includegraphics[angle=90,width=.45\linewidth]{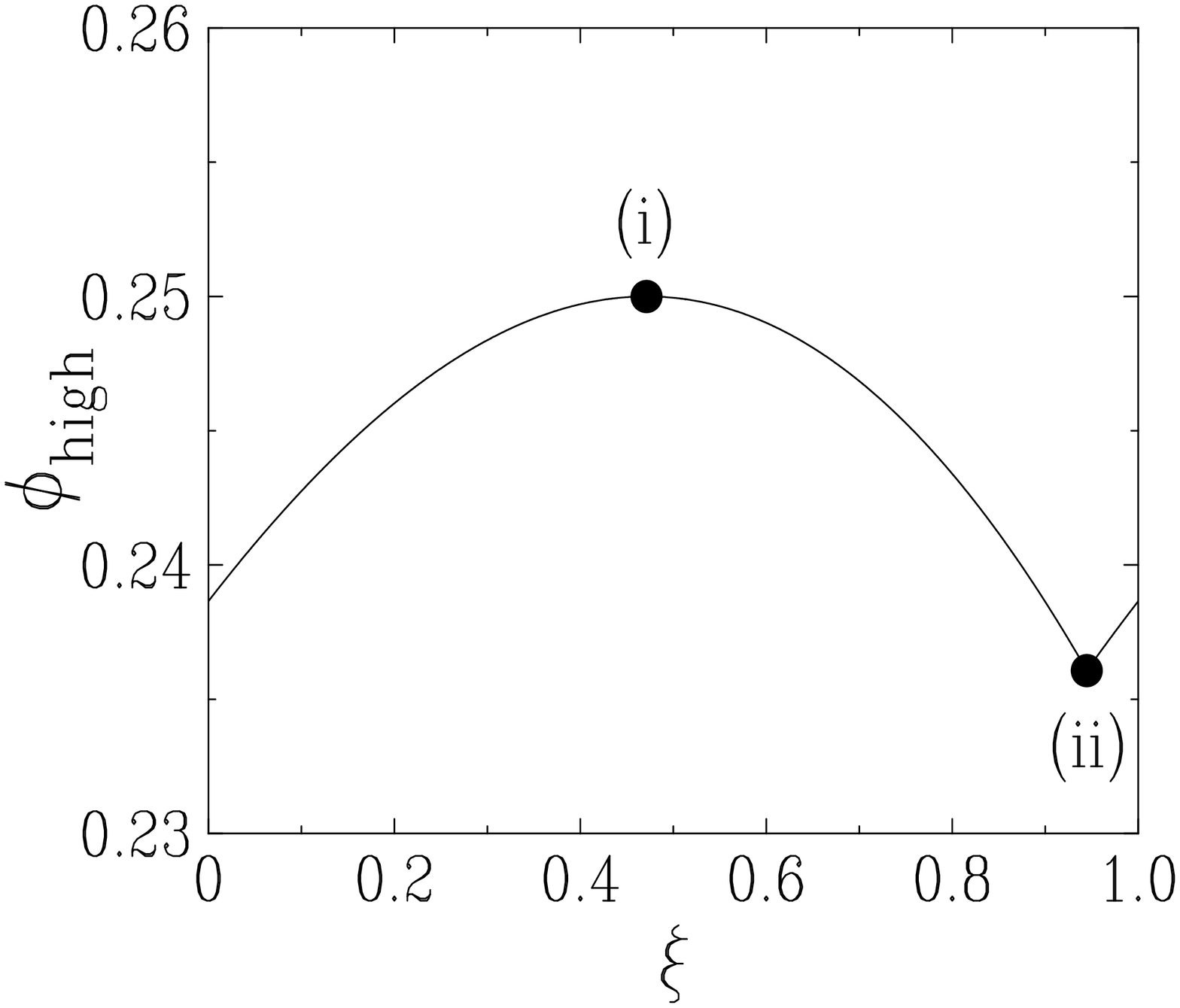}
\caption{\label{figphigh}
Plot of the highest probability $\phi_\high(\xi)$
of the largest degree against $\xi$ for geometric integer variables
with $a=1/2$.
Symbols: maximum and minimum values $\phi_{\high,\max}$ and $\phi_{\high,\min}$,
respectively corresponding to situations (i) and (ii) described in the text.}
\end{center}
\end{figure}

This phenomenon is best illustrated by considering
the highest probability $\phi_\high$,
i.e., the probability of the most probable integer $m$.
This quantity is again a periodic function of $\xi$, with unit period.
It can be checked that the highest probability is reached for
$m=1$ (i.e., $\K(n)=j+1$) if $0<\xi<\xic$
and for $m=2$ (i.e., $\K(n)=j+2$) if $\xic<\xi<1$,
where the threshold value $\xic$ depends on $a$.
Situation (ii) corresponds to $\xi=\xic$,
where $\phi_\high(\xi)$ takes its minimal value $\phi_{\high,\min}$,
the two equally probable integers being $j+1$ and $j+2$.
Situation (i), where $\phi_\high(\xi)$ takes its maximal value
$\phi_{\high,\max}$, takes place near $\xi\approx\xic-1/2$.
Figure~\ref{figphigh} shows a plot of $\phi_\high$ against $\xi$ for $a=1/2$.
The minimum $\phi_{\high,\min}=\sqrt5-2\approx0.236067$
is reached at the threshold $\xic\approx0.944744$,
whereas the maximum $\phi_{\high,\max}=1/4$ is reached for $\xi\approx0.471233$.

\subsection*{A2.~Distribution of the number of co-leaders}

The distribution of the number of co-leaders $C(n)$
can be obtained as follows.
The expansion~(\ref{phiserie}) allows one
to readily write down the joint probability of $\K(n)$ and~$C(n)$~as
\beq
\prob{\K(n)=k,C(n)=C}=\bin{n}{C}f_k^C(1-F_k)^{n-C},
\label{joint}
\eeq
so that the distribution $\rho_C(C)=\prob{C(n)=C}$ of the number of co-leaders
reads
\beq
\rho_C(C)=\bin{n}{C}\sum_{k=1}^\infty f_k^C(1-F_k)^{n-C}.
\eeq
Using~(\ref{fdef}),~(\ref{Fgeo}),
the `exponentiation' valid at large $n$,
and the definitions~(\ref{jdef}), (\ref{elldef}),
this expression can be recast as
\beq
\rho_C(C)\approx\frac{(1-a)^C}{C!}
\sum_{m=-\infty}^\infty a^{(m-1-\xi)C}\,\e^{-a^{m-1-\xi}}.
\eeq
The sum again defines a periodic function of $\xi$,
depending on the value of $C$, with unit period.
Neglecting tiny periodic oscillations,
we are left with the following asymptotic distribution
of the number of co-leaders:
\beq
\rho_C(C)=\frac{1}{\abs{\ln a}}\,\frac{(1-a)^C}{C},
\eeq
known as a {\it logarithmic distribution}.
In particular, the asymptotic mean number of co-leaders reads
\beq
\mean{C}=\frac{1-a}{a\abs{\ln a}}.
\label{meanC}
\eeq

\end{document}